\documentclass[usenatbib]{aa}
\usepackage{natbib}
\usepackage{xcolor}
\usepackage{graphicx}	
\usepackage{amsmath}	
\usepackage{hyperref}
\usepackage{txfonts}
\usepackage{multicol} 
\usepackage{tikz}

\usepackage{graphicx}
\newcommand\ourobject{{\it Alaknanda }}
\newcommand\ourobjectnospace{{\it Alaknanda}}

\renewcommand{\linenumbers}{}
\nolinenumbers 

\begin{document}

\title{A grand-design spiral galaxy 1.5 billion years after the Big Bang with JWST}

\author{Rashi Jain\inst{1}\fnmsep\thanks{e-mail:rjain@ncra.tifr.res.in}
 and Yogesh Wadadekar\inst{1}}
\institute{National Centre for Radio Astrophysics - Tata Institute of Fundamental Research, Post Bag 3, Ganeshkhind, Pune 411007, India}
\date{}
\abstract{We report the discovery of \ourobjectnospace, a large ($\sim10$ kpc diameter), massive ($\log(M_\star/M_\odot)\sim10.2$), candidate grand-design spiral galaxy with photometric redshift $z_{phot}\sim4.05$ in the UNCOVER and Medium band, Mega Science surveys with JWST. This is among the highest redshift spiral galaxies discovered with JWST. Our morphological analysis using GALFIT reveals that this galaxy is a well-formed disk, with two symmetric spiral arms that are clearly visible in the GALFIT residual. In the rest-frame near-UV and far-UV, we clearly see the beads-on-a-string pattern of star formation; in the rest-frame visible bands, each string appears as an arm. Spectral energy distribution modeling using the BAGPIPES and Prospector codes is strongly constrained by detections and flux measurements in 21 JWST and HST filters. From the BAGPIPES modeling, the stellar mass-weighted age is $\sim 199$ Myr, implying 50\% of the stars in the galaxy formed after $z\sim4.6$. This is a highly star-forming galaxy with a star formation rate (SFR) of $\sim 63 \, M_\odot \, \text{yr}^{-1}$. We detect flux excesses in the F250M and F335M filters due to the presence of H-$\alpha$+[NII] and [OIII]+H-$\beta$ emission line complexes respectively. Detection of a spiral galaxy at $z \sim 4$ indicates that massive and large spiral galaxies and disks were already in place merely 1.5 billion years after the Big Bang. Future observations with NIRSpec IFU and ALMA will be able to probe the kinematics of the galactic disk, throwing light on the possible origin of the spiral arms in this galaxy.}


\maketitle

\section{Introduction} \label{sec:intro}

Spiral arms are the primary distinguishing feature of the majority of disk galaxies in the nearby Universe. Spiral galaxies come in different flavors, ranging from grand-design and multi-arm spiral galaxies with prominent arms to flocculent spiral galaxies having rather small or patchy arms \citep{Elmegreen_2014}. Two of the most popular theories for the origin, longevity and structure of spiral arms are the Quasi-Steady-State Density Wave theory \citep{Lin_Shu_1964} and swing amplification of local gravitational instabilities \citep{Toomre_1977,Toomre_1981}.

It is unclear when and how spiral galaxies first emerged in the early Universe. Using observations with the Hubble Space Telescope (HST), it seemed that disk galaxies, and thus by extension spiral galaxies, become increasingly rare in the high-redshift Universe. The rarity of high-redshift spirals might be caused by galaxies being dynamically hot at those early epochs. Dynamically hot systems tend to form clumpy structures, whereas cold disks with low-velocity dispersion are required to form spiral arms \citep{Genzel_2006,Elmegreen_2014}. Additionally, all spiral galaxies may not have formed via similar processes; there might exist parallel mechanisms such as bar-driven \citep{Elmegreen_1982} and tidally induced \citep{Toomre_1972} spiral structure formation. The observed scarcity of spiral galaxies in the early Universe may also be a consequence of the predominance of spiral arm formation mechanisms that only form short-lived spiral patterns, at those early epochs. Cosmological dimming, morphological k-correction \citep{Rawat_2009} and observational constraints may also play a important role in the observed rarity of high redshift spirals. As redshift increases, it becomes increasingly difficult to identify and study the spiral pattern in galactic disks due to these limitations of sensitivity and resolution. Even with the deepest HST observations, studies of spiral galaxies were limited to $z< 3$. A study of spiral galaxies in the CANDELS field by \citet{Margalef_Bentabol_2022} suggests that the spiral galaxy fraction decreases monotonically with redshift. Using HST data, only two spiral galaxies have been discovered at a redshift of $z > 2$ \citep{Law_2012,Yuan_2017}.

James Webb Space Telescope (JWST)'s unprecedented sensitivity and resolution in the observed frame near-infrared (NIR) \citep{Gardner_2023} that provides access to the optical rest frame at $z>3$, has made it possible to detect spiral galaxies at these redshifts, for the first time. With JWST, astronomers have discovered a significant population of morphological disk galaxies at high redshifts ($ z > 3 $) \citep{Kartaltepe_2023,Ferreira_2022,Ferreira_2023,Robertson_2023,Nelson_2023,Jacobs_2023}. The discovery of high redshift disk galaxies poses challenges to our current understanding of galaxy formation \citep{Kartaltepe_2023,Ferreira_2022}. JWST's sensitivity also allows us to detect spiral features in high-redshift disk galaxies. Recently, using JWST observations, a few studies such as \citet{Wu_2022}, \citet{Fudamoto_2022}, \citet{Costantin_2023} and \citet{Wang_2025} have discovered individual spiral galaxies at high redshifts of 3.06, 2.463, 3.03, and 3.25, respectively. \citet{Xiao_2024} have also reported the discovery of a candidate spiral galaxy at a redshift of 5.2. Another study by \citet{Kuhn_2024} finds that the observed spiral galaxy fraction decreases from 46\% at $z\sim1$ to 4\% at the redshifts of $z \sim 3$. Using an extrapolation method of artificially redshifting spiral galaxies to high redshifts, \citet{Kuhn_2024} claim that the intrinsic fraction of spiral galaxies should be $\sim$25\% at the redshifts of $z \sim3$. The mismatch between the observed fraction and the prediction likely arises due to observational biases in detection. Observationally, it is crucial to discover as many spiral galaxies at high redshifts as possible, to enable statistical studies of their properties. Even with JWST observations, only a few individual spiral galaxies have been found at $z \gtrsim 3$, although the number of non-spiral disk galaxies discovered is considerably larger.

In a few complementary studies, astronomers have also discovered spiral structures in galactic disks at comparably high redshifts through ground-based observations using ALMA \citep{Tsukui_2021,Tsukui_2024,Huang_2023}. The study by \citep{Tsukui_2021,Tsukui_2024} has reported the discovery of spiral structure at a high redshift of $ z\sim4$ through submm ALMA observations. However, the HST image of this ALMA detected galaxy does not show a spiral structure because its observed frame optical light is dominated by the quasar at the center of the galaxy\citep{Tsukui_2023b}. Due to this, it is not possible to confirm whether the spiral arms in this ALMA-detected galaxy are stellar arms, or just spiral structures in the gas.

In this paper, we present the discovery, in the Abell 2744 cluster field, of a candidate two-armed, grand-design spiral galaxy, with a redshift of $z\sim4$,  when the Universe was only $\sim1.5$ billion years old. We name this galaxy as \ourobject - a Himalayan river, one of the headstreams of the {\it Ganga} and sister river of the {\it Mandakini}, which, in turn, lends its name to the Milky Way galaxy. \ourobject is a massive (stellar mass $\sim$10$^{10.2}$\(M_\odot\)) and highly star forming galaxy with a star formation rate (SFR) of $\sim$63 \(M_\odot\) yr$^{-1}$ making it analogous to star-forming spiral galaxies in the nearby Universe. It is one of the most distant spiral galaxies discovered by JWST.

In Section~\ref{sec:dataset}, we describe the UNCOVER dataset. In Section~\ref{sec:methods} we discuss our methods for SED modeling and quantitative parametric and non-parametric measures of galaxy morphology. In Section~\ref{sec:results}, we present our results and interpret the possible consequences of this discovery on models of high redshift spiral galaxy formation. We conclude and list the open questions in Section~\ref{sec:summary}.   Throughout, we assume a standard cosmological model with $\Omega_M = 0.3$, $\Omega_\Lambda = 0.7$ and Hubble’s constant $H_0 = 70$ km s$^{-1}$ Mpc$^{-1}$.

\section{Dataset} 
\label{sec:dataset}

We utilise the publicly available JWST and HST data from the Ultradeep NIRSpec and NIRCam Observations before the Epoch of Reionization (UNCOVER) survey Data Release 3 and 4 \citep{Bezanson_2022,Weaver_2024,Suess_2024} in the Abell 2744 field. The UNCOVER DR3 dataset includes photometry and deep mosaics in a total of 15 JWST/NIRCam \citep{Rieke_2023} (F090W, F115W, F150W, F200W, F277W, F356W, F410M, F444W), HST/ACS (F435W, F606W, F814W), and HST/WFC3 (F105W, F125W, F140W, F160W) filters. It also includes observations from the Medium bands Mega Science survey in 11 additional medium band filters (F140M, F162M, F182M, F210M, F250M, F300M, F335M, F360M, F430M, F460M, F480M) and the two shortest wavelength broad-band filters (F070W and F090W). We note that the overlap in the footprints among the filters listed is considerable, but not complete. The addition of data in a large set of medium-band filters through the Mega Science observations enables the correct disentangling of the relative contribution of line and continuum emission to broad-band photometry, leading to non-degenerate photometric redshifts and accurate stellar population synthesis modeling. The UNCOVER DR3 dataset provides aperture photometry for over 70,000 objects over an area of 56.2 arcmin$^{2}$. The UNCOVER DR3 SUPER catalog provides photometry for each object with an appropriate aperture size. Over the limited area of overlap between the UNCOVER and Medium Band Mega Science surveys, the UNCOVER DR3 datasets include observations in all NIRCam medium and broad-band filters.  The DR3 dataset includes image mosaics that are astrometrically matched and pixel aligned with a pixel scale of 0.02 arcsec/pixel in the short wavelength filters (with wavelength less than that of F250W) and 0.04 arcsec/pixel in the long wavelength (F250W and longer wavelength) filters. As value additions, the DR4 release provides photometric redshifts derived via stellar population synthesis (SPS) codes Prospector \citep{Johnson_2021,Wang_2024} and EAZY \citep{Brammer_2008}. Additionally, DR4 includes an SPS catalog that contains parameters such as stellar mass, star formation rate (SFR), metallicity and dust extinction derived via Prospector. Abell 2744 is a massive galaxy cluster that magnifies the background galaxies up to dozens of times through gravitational lensing. The DR4 catalog also supplies us with the strong lensing magnification of each galaxy derived via an updated version of the mass model described in \citet{Furtak_2023}. In this work, we perform SED modeling using BAGPIPES by utilizing the medium and broadband photometry from the UNCOVER DR3 SUPER catalog that provides the photometry in a 1.4" aperture for \ourobject given the extended nature of the source. The shortest wavelength filter we use in our analysis, is F606W and the longest is F480M.

\begin{table*}
\caption{Basic properties of the \ourobject galaxy }
\label{tab:observes_prop}
\small
\centering
\begin{tabular}{lccccccccccccc}
\hline\hline
ID&RA&Dec&z\_eazy&z\_eazy\_16 &z\_eazy\_50 &z\_eazy\_84 & z\_p\_16 & z\_p\_50 & z\_p\_84 & R\_h & R\_c & mu\\

&(deg)&(deg)&&z & & &  & &  & (kpc) & (kpc) & \\
\hline
42812 & 3.5237 & -30.37241 & 3.90 &3.86 &3.89 &3.92 &3.88 &3.91 &3.95 &3.03 &8.14 &2.273 \\

\hline
\end{tabular}
\tablefoot{
The data presented in this table for \ourobject are taken from the UNCOVER DR3 and DR4 catalogs. The ID is the catalog ID of \ourobject from the UNCOVER DR3 catalog. z\_eazy is the maximum likelihood photometric redshift derived via EAZY. z\_eazy\_16/50/84 and z\_pros\_16/50/84 are 16th/50th and 84th percentiles of EAZY and Prospector output photometric redshift posterior respectively. R\_h and R\_c represent the half-light radius and circularised Kron radius \citep{Kron_1980} of the galaxy. mu is the strong lensing magnification.}
\end{table*}

\section{Methods \label{sec:methods}}
\subsection{Identification}

 \ourobject was first recognized as an unusual galaxy in a larger study (Jain et al. 2025 in prep.) of morphological properties of high redshift galaxies at $3 <z < 6$ using UNCOVER data in the A2744 field. In this study, we have conducted a visual and quantitative morphological classification of galaxies and identified \ourobject as a spheroid + disk galaxy i.e. this galaxy contains a central bulge component and a disk. \ourobject is the only galaxy that shows a grand-design spiral pattern with two well-formed spiral arms (UNCOVER DR3 SUPER catalog id of 42812) in the redshift range of our sample. We have listed some of the relevant parameters such as sizes and strong lensing magnification from the UNCOVER DR3 and DR4 catalogs for \ourobject in Table~\ref{tab:observes_prop}.
 
 From the UNCOVER DR3 and DR4 dataset, we learn that the maximum likelihood photometric redshift of \ourobject derived via two independent SED modeling codes Prospector and EAZY is $3.90$ and $3.89$, respectively. These broadly consistent photometric redshifts likely arise because this galaxy is detected, at high significance, in all JWST/ NIRCam broadband and medium band filters. It is not well detected in any of the HST/ACS filters where the footprint covers this galaxy, due to the much shallower imaging with the HST. In the F606W HST/ACS filter, the aperture photometry indicates a marginal $3\sigma$ detection. It is not covered in the footprint of any of the HST/WFC3 filters. We use data in all JWST/NIRCam filters and the HST/ACS F606W filter. The photometric redshifts obtained with Prospector and EAZY have narrow posteriors with the values of 16th, 50th and 84th percentiles of photometric redshift from both the codes listed in Table~\ref{tab:observes_prop}.

There also exists a small spheroid galaxy, near the southern end of the galactic disk. This spheroid galaxy has an UNCOVER DR3 catalog id of 42811 (see the last figure in the bottom panel of Fig.~\ref{fig:SP31564}). In the UNCOVER DR4 catalog its photometric redshift obtained with Prospector is $z\sim3.85$. This galaxy also has a robust NIRCam/GRISM spectroscopic redshift of 3.973671 from the All the Little Things (ALT) survey \citep{Naidu_2024} measured using two different emission lines which makes it a candidate satellite galaxy of \ourobjectnospace, rather than a chance interloper. Moreover, this spheroid appears to be in the foreground of the disk of \ourobject further indicating that \ourobject is indeed at a redshift, at least as high, as the spectroscopic redshift of the spheroid galaxy.

\subsection{SPS modeling using BAGPIPES}

To further verify the photometric redshift of \ourobjectnospace, and to measure its properties using the stellar population synthesis (SPS) technique, we carry out spectral energy distribution (SED) modeling with another independent state-of-the-art modeling code, Bayesian Analysis of Galaxies for Physical Inference and Parameter EStimation (BAGPIPES) \citep{Carnall_2018}. BAGPIPES provides a framework that allows us to model the observed SEDs of galaxies by generating synthetic model spectra based on the input parameters. BAGPIPES implements SPS models from the 2016 version of the \citet{Bruzal&Charlot_2003} models and an initial mass function (IMF) by \citet{Kroupa_2002}. BAGPIPES allows us to provide the range and priors for the input parameters and set different parametric star formation histories. It utilizes a nested sampling algorithm PyMultinest \citep{Feroz&Hobson_2008,Feroz_2009, Feroz_2019,Buchner_2014} to derive samples from input parameters to generate model SEDs. By sampling a large number of model SEDs, BAGPIPES derives joint constraints on photometric redshifts and SPS parameters such as stellar mass, star formation rate (SFR), metallicity, and dust extinction.

We fit two families of star formation histories using BAGPIPES -- exponentially decaying (e$^{-t/\tau}$) and delayed exponentially decaying star formation (te$^{-t/\tau}$)  -- to the observed SED,  where $\tau$ is the timescale of the decrease of the star formation rate. For our modeling, we utilize the aperture photometry (corrected for the strong lensing magnification from the UNCOVER DR3 catalog) of \ourobject in all JWST/ NIRCam medium and wide-band filters and the F606W filter of HST/ACS from the UNCOVER DR3 catalog. In our SED modeling, we include the flux in the F606W HST/ACS filter despite its rather weak $3\sigma$ detection along with 20 JWST/NIRCam filters because F606W lies blueward of the Lyman-$\alpha$ emission line at $z\sim 4$ providing important constraints on the photometric redshift. We use a noise floor of 5 percent to account for additional uncertainties with the calibration of JWST data that we add in quadrature with the UNCOVER DR3 catalog flux uncertainties to calculate flux errors, as suggested by  \citep{Boyer_2022}. The details of the SED modeling and the list of input parameters provided to BAGPIPES are given in Appendix~\ref{sec:appendix}, which may be used to reproduce our results.

\subsection{Parametric and non-parametric morphology} \label{sec:parametric}

Parametric modeling of galaxies (often referred to as bulge-disk decomposition) is performed by assuming an analytic expression for the radial light profile for the main galaxy components. The bulge is usually modeled as a S\'ersic function \citep{Sersic_1968} and the disk is usually modeled as a radially decaying exponential \citep{Freeman_1970}. Our main goal in carrying out this decomposition is to accurately model the light of the disk, so that it can be subtracted out to better reveal the spiral arms, embedded within it. The most popular tool to perform a two-dimensional, two-component bulge-disk decomposition of galaxies is GALFIT \citep{Peng_2002,Peng_2010}. We ran GALFIT separately on images in 4 broadband filters (F200W, F277W, F356W, F444W). GALFIT was configured to fit the bulge with a S\'ersic light profile (with S\'ersic index ($n$) kept free) and an exponential for the disk light profile. While running GALFIT, we masked out the spheroid galaxy in the southern part of \ourobjectnospace. 

\begin{figure*}
  \centering
  \includegraphics[width = 0.85\textwidth]{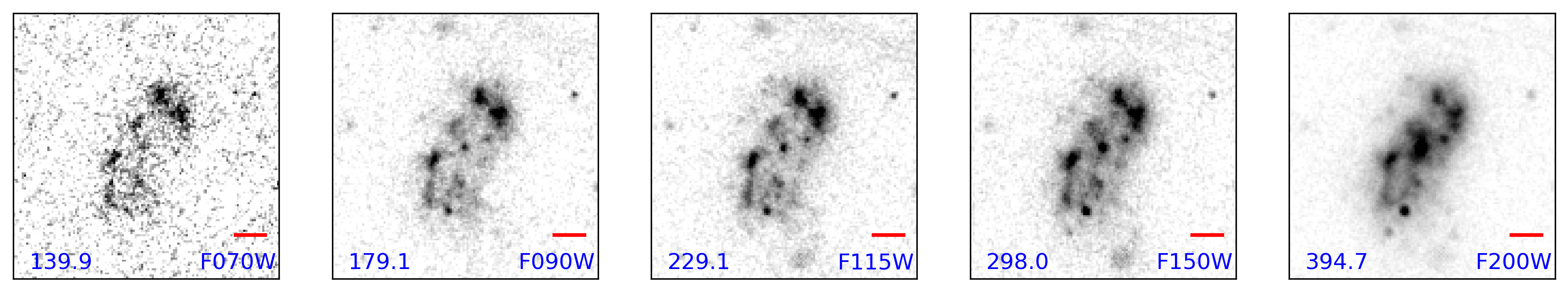}
  \includegraphics[width = 0.85\textwidth]{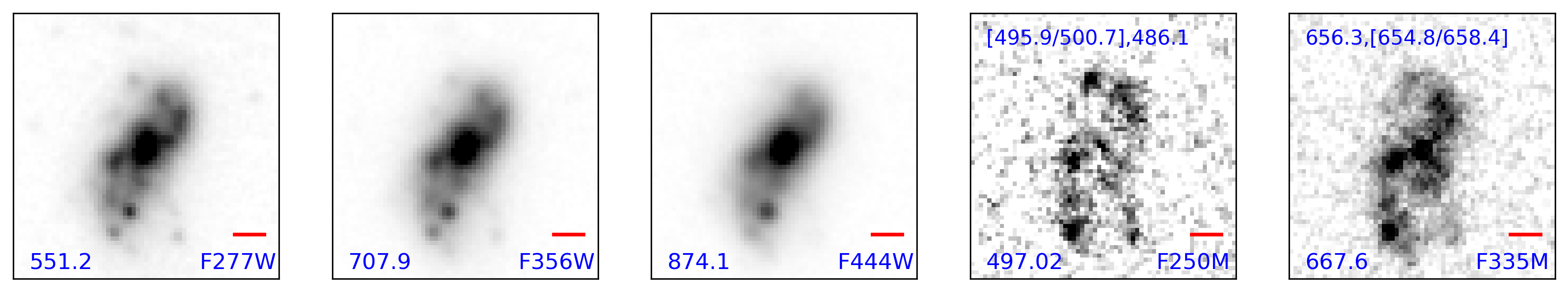}
  \includegraphics[width = 0.15\textwidth]{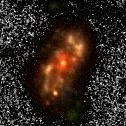}
  \includegraphics[width = 0.15\textwidth]{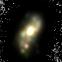}
  \includegraphics[width = 0.15\textwidth]{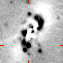}
  \includegraphics[width = 0.15\textwidth]{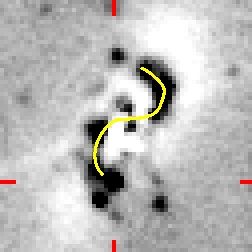}
  \caption{We show the grayscale cutouts of \ourobject in all JWST/NIRCam broadband filters in the top two panels. The red horizontal bar at the bottom right of each cutout shows a 2 kpc scale at the redshift of the galaxy. We also provide the central rest frame wavelength (in nm) for each filter at the bottom left of each cutout. The F250M and F335M filters contain the [OIII]+H-$\beta$ and H-$\alpha$+[NII] emission line complexes respectively. In the last two columns of the middle row, we show the continuum subtracted images in the F250M and F335M filters that show the regions with line emission in the disk. In these two panels, we show the rest frame wavelengths (in nm) of the [OIII] doublet and H-$\beta$ in the F250M cutout and H-$\alpha$ and [NII] doublet in the F335M cutout, at the top. The bottom row shows RGB composite images of the galaxy in sequence: RGB composite with F200W, F150W, F090W SW filters and RGB image with F444W, F356W and F277W LW filters. The last two images in the bottom row show the GALFIT residual in F277W filter obtained after subtracting the bulge + disk model derived by GALFIT and the outline of the spiral arms of the galaxy (yellow curve) superposed on the GALFIT residual. The red ticks on the x and y axes point to the location of the possible satellite spheroid galaxy at $z=3.973671$. All cutouts are $2.5 \times 2.5$ arcsec in size and North is on top and East is to the left.}
  \label{fig:SP31564}
\end{figure*}

We present the GALFIT residual in the F277W band in the bottom row of Fig~\ref{fig:SP31564}. By modeling and subtracting the light profile of the dominant disk and bulge components, the GALFIT residual allows us to clearly detect the spiral arms. Amongst its outputs, GALFIT provides the integrated AB magnitudes for the bulge and disk based on the best-fit analytic profile. We convert these magnitudes to fluxes and calculate the bulge-to-total $B/T$ luminosity ratio.

\begin{figure*}
  \centering
  \includegraphics[width=0.4\textwidth]{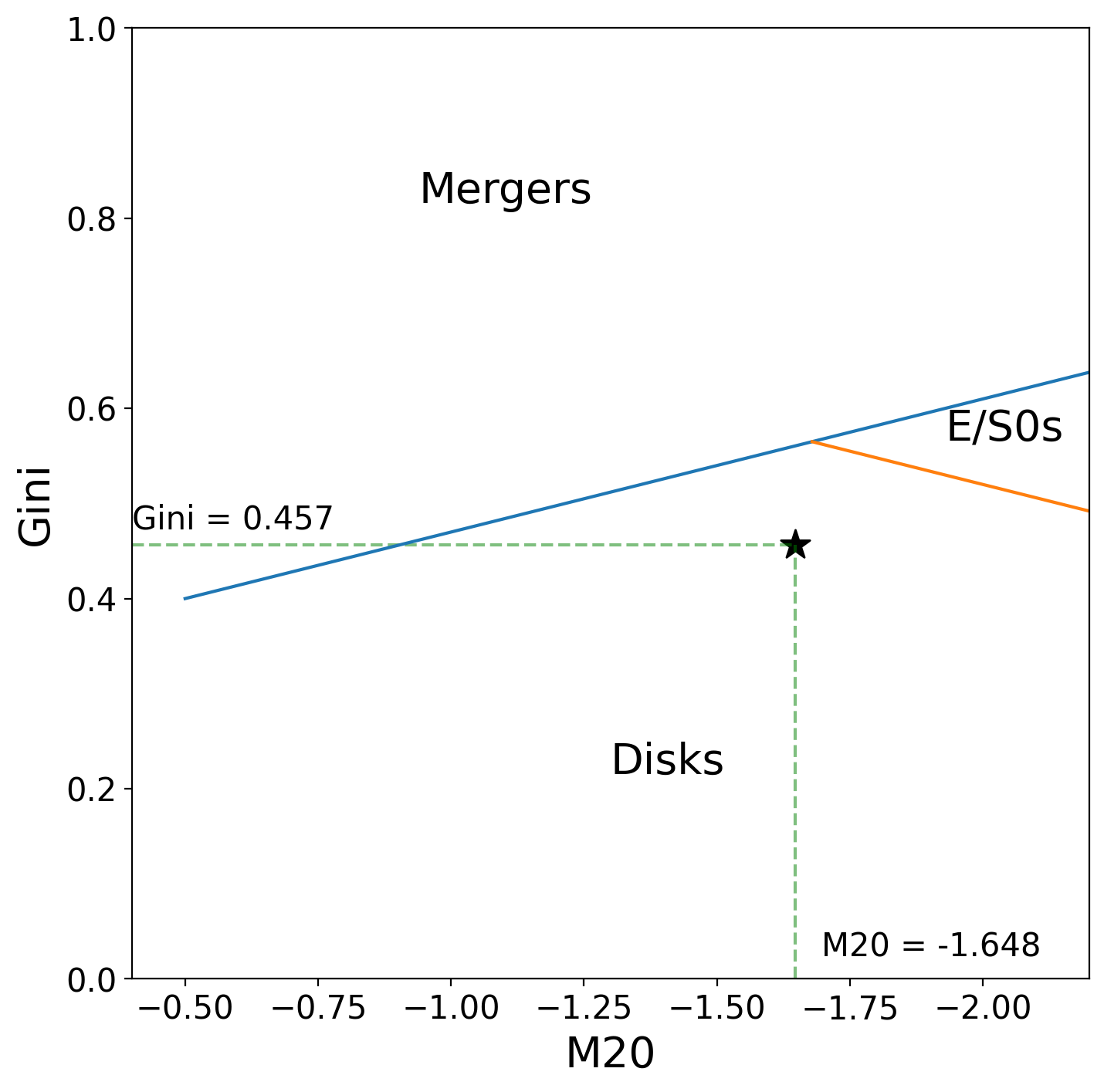}
  \includegraphics[width=0.4\textwidth]{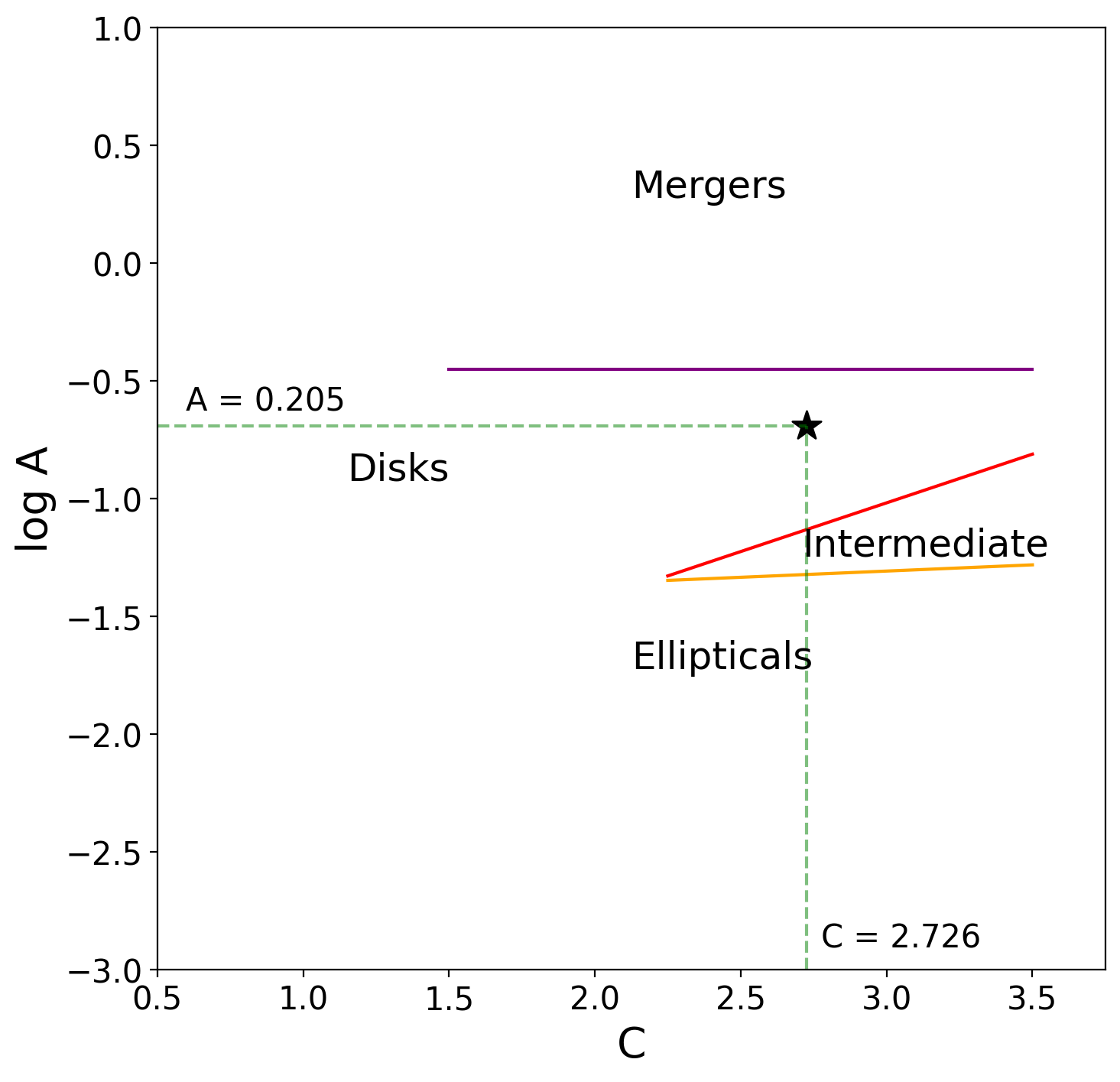}  
  \caption{The left panel shows a plot of the Gini coefficient against the M20 parameter. The blue and orange lines separating merger, disk, and E/S0 galaxies in the left panel have been taken from \citet{Lotz_2008b}. The right panel shows the log of the asymmetry against concentration. The magenta line separating mergers and disks has been taken from \citet{Conselice_2003}, where A = 0.35. The other two lines in red and orange color separating the population of disk, intermediate, and elliptical galaxies have been taken from \citet{Bershady_2000}. Exact values of the parameters for our galaxy (denoted as star symbol) are marked in both panels.}
  \label{fig:CASGM}
\end{figure*}

The morphological nature of high-redshift galaxies can also be quantified using standard non-parametric measurements. Non-parametric measurements do not assume any analytical profile and thus are well suited for high redshift galaxies, where parametric measurements are difficult due to their small angular sizes\citep{Lotz_2004}. We calculate concentration (C), asymmetry (A), \citep{Abraham_1996,Conselice_2003}, Gini coefficient (G) \citep{Abraham_2003}, and M20 \citep{Lotz_2004} (CASGM) parameters for \ourobject using the  Statmorph \citep{Rodriguez_Gomez_2019} software. As inputs, we provide the F444W filter image of \ourobject along with the segmentation map and inverse variance weight map in this filter. Disk galaxies typically have lower values of concentration and asymmetry as their light is distributed relatively symmetrically throughout the galactic disk. Gini coefficient and M20 are sensitive to the spatial distribution of light in the galaxy. The Gini coefficient ranges from 0 to 1 with its value being close to zero if the light is evenly distributed in all the image pixels that contain the galaxy's light. We make diagnostic plots of $\log A$ versus concentration (C) and Gini coefficient (G) versus the M20 parameter in Fig~\ref{fig:CASGM}. The diagnostic lines from \citet{Bershady_2000, Lotz_2004, Lotz_2008b, Conselice_2003} divide each plot into different regions that allow us to separate the galaxies into different categories like disk, ellipticals and merger/peculiar. \ourobject lies well within the disk region in both the plots, which is a pointer to its shallow disk-like profile, which is to be expected for a grand design spiral galaxy.

We show ${2.5\times 2.5}$ arcsec cutouts of \ourobject in all broadband JWST/NIRCam filters in Fig~\ref{fig:SP31564}. We have also produced RGB images of this galaxy using filters from the SW and LW channels of the JWST/NIRCam using the software GnuAstro \citep{gnuastro,makecatalog,noisechisel_segment_2019} ( bottom row of Fig~\ref{fig:SP31564}). For SW filters, we use F090W, F150W, and F200W filters and for LW filters, we use F277W, F356W, and F444W filters as our blue, green, and red channels, respectively. We create difference images in F335M and F250M filters that contain H-$\alpha$+[NII] and [OIII]+H-$\beta$ emission line complexes respectively. We calculate and subtract the continuum flux from F335M and F250M filters by taking a weighted average of the observed flux in adjacent medium/broad band filters \citep{Tacchella_2024}. For F335M, we use F330M and F360M while for F250M, we use F210M and F277W filters. The continuum subtracted emission line images are shown in the middle panel of Fig~\ref{fig:SP31564}.

\section{Results and discussion} \label{sec:results}
\subsection{Robust photo-z and SPS outputs from SED modeling}

Our SED modeling using BAGPIPES yields an output photometric redshift of the galaxy to be $4.045_{-0.034}^{+0.034}$ with tight constraints on the output photo-$z$ posterior (see Fig.~\ref{fig:Bagpipes_fit_1dpost}). The best-fit SED model for delayed exponential SFH passes through almost all the data points within the error bar. We note that the error bar for the shortest wavelength data point (F606W) is quite large due to the poor sensitivity of the HST observations. The tight constraints on the position of the Lyman and Balmer breaks and the strong emission lines provided by the numerous, closely-spaced JWST/NIRCam medium-band filters become apparent. The presence of several emission lines is as expected for a star-forming galaxy. The stellar mass, SFR and the logarithm of the specific star formation rate (sSFR; yr$^{-1}$) are 10$^{10.17}$ \(M_\odot\), 62.75 \(M_\odot\) yr$^{-1}$ and, -8.37 respectively as derived from the delayed exponential SFH model. We also derive the values of mass-weighted age and dust attenuation $A_{v}$ assuming the  \citep{Calzetti_2000} dust law. The best-fit model from BAGPIPES shows the presence of the Lyman break, along with flux excess in the F250M and F335M medium band filters due to the presence of [OIII] + H-$\beta$ and H-$\alpha$+[NII] emission line complexes, respectively. The excess flux seen in these filters is consistent with \ourobject being located at $z\sim4$. The SED also shows a well-defined Balmer break in the medium band filter F182M. From the UNCOVER DR4 SPS catalog, we learn that the 16th, 50th, and 84th percentiles of output photometric redshift posterior are 3.878, 3.908 and 3.951 respectively as derived by Prospector (Table~\ref{tab:Table_sps}). As with BAGPIPES, the best-fit Prospector model also provides a good fit to the observed data points with some discrepancies between observed and model fluxes in a few filters. For example, the best-fit Prospector model underestimates the model flux as compared to the observed flux in the F250M filter that contains the [OIII]+H-$\beta$ emission line complex which BAGPIPES models correctly (Fig.~\ref{fig:Pros_best}). Additionally, the [OIII] + H-$\beta$ emission line complex will boost the filter F250M only at redshifts of $z \geq 4$, which is inconsistent with the Prospector photometric redshift solution (see Fig 6 of \citet{Suess_2024}). We summarize our results from SED modeling via BAGPIPES (using two different SFHs) and the DR4 Prospector results in Table~\ref{tab:Table_sps} for comparison. The small differences between the BAGPIPES and Prospector outputs are not unexpected because the two codes use different templates incorporating different star formation histories and distinct treatment of the nebular and dust components \citep{Pacifici_2023}. Prospector uses FSPS \citep{Conroy_2010,Conroy&Gunn_2010} templates while BAGPIPES uses BC03 (2016 version) \citep{Bruzal&Charlot_2003} templates. We choose to use BAGPIPES SPS outputs for further analysis as the best fit BAGPIPES SED model is completely consistent with enhanced flux in two medium-band filters due to the H-$\alpha$+[NII] and [OIII]+\, H-$\beta$ emission line complexes. We note that the key result of this paper -- the discovery of a candidate grand design spiral galaxy at $z\sim4$ -- is not at all changed by the choice of SPS code. We have also performed SED modeling for the spheroid candidate satellite galaxy (UNCOVER DR3 ID 42811) using BAGPIPES and derive a photometric redshift of 3.987, which is in excellent agreement with the spectroscopic redshift (3.973671) of this galaxy. 

\begin{table*}
\caption{Stellar population synthesis results from BAGPIPES and Prospector}
\label{tab:Table_sps}
\small
\centering
\begin{tabular}{lccc}
\hline\hline
Parameter & BAGPIPES(delayed) & BAGPIPES(exp) & Prospector\\

\hline
z\_ml & 4.035 & 4.060 & 3.885 \\
z\_p & 4.011 / 4.045 / 4.079 & 4.013 / 4.048 / 4.080 & 3.878 / 3.908 / 3.951 \\
mass\_p  ($\log(M_*/M_\odot$)) & 10.14 / 10.17 / 10.21 & 10.12 / 10.16 / 10.20 & 9.81 / 9.93 / 10.03 \\
SFR\_p (\(M_\odot\) yr$^{-1}$) & 59.27 / 62.75 / 70.58  & 53.58 / 63.25 / 73.91 &  30.23 / 36.63 /, 43.37\\
log(sSFR\_p (yr$^{-1})$) & -8.48 / -8.37 / -8.29 & -8.47 / -8.36 / -8.25 &  -8.54 / -8.35 / -8.19 \\
A$_{v}$\_p & 0.84 / 0.89 / 0.94 &0.84 / 0.89 / 0.94  & \\
dust2 ($\tau$) & & & 0.34 / 0.40 / 0.47\\
MWA\_p (Gyr) & 0.159 / 0.199 / 0.265 &  0.121 / 0.157 / 0.205  & 0.176 / 0.316 / 0.491 \\

\hline
\end{tabular}
\tablefoot{
This table shows the SPS properties of \ourobject derived from fitting two SFHs from BAGPIPES (delayed exponentially decaying SFH and exponentially decaying SFH) and Prospector SPS values from UNCOVER DR4. z\_ml is the maximum likelihood redshift obtained by SED modeling. In other rows, we show, in order, the 16th/50th/84th output posterior percentiles of photometric redshift, stellar mass, SFR, logarithm of sSFR, dust extinction (A$_{v}$), optical depth ($\tau$) and mass-weighted age (MWA). A$_{v}$ and dust2 are BAGPIPES and Prospector outputs, respectively and both are measures of the dust content where A$_{v}$ is the absolute attenuation in the V-band and dust2 ($\tau$) is the optical depth of the diffuse dust.}

\end{table*}

\subsection{Morphology and its correlation with global physical SPS properties}
\label{subsec:morph_sps}
\ourobject is a well-formed disk galaxy with a prominent spiral structure at a high redshift of $z \sim 4$. It shows a two-armed spiral structure with a bright central bulge and a large, extended disk. We visually measure the angular extent of \ourobject to be $\sim 1.5$ arcsec in the F444W filter which is our longest wavelength broadband filter. This angular size corresponds to a linear size of $\sim 10$ kpc at the redshift of 4.05 for our chosen cosmology. The spiral structure is visible in all JWST/NIRCam broadband and medium-band filters. In the rest frame UV, the star-forming regions in each arm display the familiar beads-on-a-string type structure \citep{Elmegreen_2014}, commonly seen in nearby spirals. In the rest frame visible bands, each string appears as an arm, and the star-forming regions progressively lose their prominence as we move to longer wavelength filters. From our GALFIT modeling (see Section~\ref{sec:parametric}) in the F277W filter, the dominant component is the exponential disk with a best fit scale length of 0.29$\pm$0.00 arcsec ($\sim2.00$ kpc). The S\'ersic model for the bulge has a S\'ersic index of ~2.00$\pm$0.22 with effective radius of 0.32$\pm$0.04 arcsec (2.21$\pm$0.28 kpc). The integrated AB magnitudes for the disk and bulge component are 23.00$\pm$0.01 and 24.93$\pm$0.09 in the F277W filter. We convert these magnitudes to fluxes and derive the bulge to total luminosity ($B/T$) ratio which is 0.14 in the F277W filter. We observe that $B/T$ increases monotonically from the F200W filter towards the longer wavelength filters. In the longest wavelength broadband filter (F444W) we derive a $B/T$ ratio of $\sim0.18$ indicating that this galaxy is an overwhelmingly disk-dominated system in the rest frame visible band from $\sim400$ nm to $\sim880$ nm. The diagnostic plots of $\log A$ vs concentration and Gini coefficient vs M20 from our non-parametric quantitative morphological analysis also evidently confirm the disk morphology of \ourobject (Fig~\ref{fig:CASGM}).

Our SPS modeling with BAGPIPES reveals that \ourobject is a massive 
(stellar mass $\sim$10$^{10.2}$\(M_\odot\)) and vigorously star-forming galaxy with SFR  of $\sim$63 \(M_\odot\) yr$^{-1}$ and log sSFR (yr$^{-1}$) of -8.37. The large mass, high SFR and sSFR are consistent with its morphological structure which is indicative of a typical star-forming galaxy. The dust extinction of A$_v \sim 0.9$ shows that \ourobject is a moderately dusty galaxy at this high redshift of $z \sim 4$ with dust extinction significantly lower than that of dusty ULIRG's observed with JWST \citep{Huang_J_2023}. From the continuum subtracted images in F250M and F335M filters (Fig~\ref{fig:SP31564}), we see that the spiral arms show flux excess due to the presence of emission line complexes. In F335M filter, we see a faint H-$\alpha$ emission throughout the disk as well. These images indicate that the star formation is occuring throughout the disk with several bright star forming regions located in the spiral arms. We derive the mass-weighted age of this galaxy to be 199 Myr. The onset of star formation in this galaxy using a delayed exponential star-formation history model occurs approximately 914 Myr after the Big Bang (see Fig.~\ref{fig:Bagpipes_corner_sfh}). Hence, this disk galaxy has assembled a mass of 10 billion solar masses in a few hundred million years and the age of the Universe at this redshift $z\sim4$ is only about 1.5 Gyr. 

\subsection{Insights into spiral arm formation at high redshift}
Before the advent of JWST, the broad consensus in the literature was that galaxies generally have a clumpy structure at $z > 1$ \citep{Guo_2015} and stable disk galaxies are rare. Unstable primordial disks formed by smooth accretion get fragmented into clumpy disks typically around $z\sim2-3$. At later epochs ($z\sim1$), the velocity dispersion of clumpy galaxies decreases and their disks become more stable. In clumpy galaxies, the clumps interact with each other and dissolve by exchanging angular momentum, ultimately forming exponential disks. These dynamically cold disks can form spiral arms within 1 Gyr provided they do not get destroyed by major mergers \citep{Bournaud_2007,Bournaud_2009,Genzel_2006,Elmegreen_2009a,Elmegreen_2009b, Elmegreen_2014,Elmegreen_DM_2007}. This indicates that spiral galaxies may have evolved from a clumpy phase i.e., the dissolution of clumps in the high redshift clumpy galaxies may be a potential mechanism for the origin of spiral galaxies. Such a formation scenario is consistent with the first appearance of spiral galaxies between redshifts of 1.4 and 1.8 in HST observations \citep{Elmegreen_2014}. According to a more recent statistical study by \citet{Margalef_Bentabol_2022} in the CANDELS field with HST observations, spiral galaxies make their appearance between the redshifts of 1.5 to 3. Cold disks are required for density waves to emerge, which later form long-lived spiral patterns. One such example is the spiral galaxy discovered by \citet{Yuan_2017} with a cold disk at $z=2.54$. 

Interestingly,  \citet{Feng_2015} had predicted  -- using the BlueTides simulation \citep{Feng_2016} -- that rotationally supported kinematic disks should dominate the population of massive galaxies in the early universe. According to this study, these high-redshift disk galaxies should have stellar masses similar to the Milky Way but with substantially smaller sizes. In broad agreement with the predictions of this simulation, JWST has begun to unveil a substantial population of galaxies with disk morphology at $z > 3$ \citep{Kartaltepe_2023,Ferreira_2022,Nelson_2023,Robertson_2023}. \citet{Rowland_2024} have recently discovered a dynamically cold disk galaxy at $z=7.3$ with ALMA. Another study by \citet{Graaff_2024} has found dynamically cold disks in the redshift range of $5.5 < z < 7.5$; however, it is still not confirmed when exactly cold disks started to settle in the early universe \citep{Danhaive_2025}. In light of these discoveries, it has been proposed that cold disks were already present in the Universe at redshifts of $z \sim 3-4$ \citet{Kuhn_2024}. These authors find that the majority of their sample galaxies show coexistence of spiral structure and clumps which is true for \ourobject as well.  

At high redshifts, grand design spiral patterns could also form through perturbations produced by minor galaxy mergers or bar interactions \citep{Elmegreen_2014, Law_2012} in relatively unstable hot disks. The spiral structures produced by mergers are short-lived and have material arms i.e., the spiral arms that follow the kinematics of the stellar disk \citep{Oh_2008}. Tidal interactions with a companion may also induce spiral arms in a galactic disk, depending upon their relative mass and the orientation of the orbital plane with respect to the disk galaxy. \citep{Hernquist_1992,Oh_2008, Toomre_1972,Dobbs_2014,Struck_2011}. A candidate satellite spheroid galaxy in the southern part of \ourobject may be responsible for the spiral structure in this galaxy.

It is possible to form massive galaxies with an extended disk component in major galaxy mergers\citep{Springel_2005}. However, the galaxies formed via mergers have only a few percent of their mass in the disk and the majority of the mass exists in the central bulge component and their spheroidal halo \citep{Robertson_2006,Hopkins_2008}. Therefore, formation of a massive spiral disk galaxy such as \ourobject at $z\sim4$ may indicate a galaxy growth scenario by smooth accretion rather than mergers which is also consistent with its low B/T ratio (see subsection ~\ref{subsec:morph_sps}). \citep{Bournaud_2009}.

Another possibility is that stellar bars may produce perturbations in the galactic disk and provide an overall pattern speed to the disk to induce density waves \citep{Salo_2010}. \citet{Huang_2023} have also reported the discovery of a spiral galaxy at $z=2.467$ whose formation may be bar-driven. However for \ourobject from observations in the higher resolution SW filters, the presence of a relatively large bar ($\gtrapprox 2$~kpc) is ruled out, leading us to believe that a bar-driven spiral structure scenario is not plausible for this galaxy.

\section{Conclusion and the way forward}
\label{sec:summary}
\ourobject is a candidate grand design spiral galaxy at $z\sim4$ with two independent SPS codes BAGPIPES and Prospector, being in broad agreement on its photometric redshift. This discovery raises several questions. What is the dominant mechanism responsible for the origin of spiral structure in this galaxy? Is it induced by the dissolution of clumps over time or via interaction with its candidate satellite galaxy? How did \ourobject acquire such a large disk in such a short time? When did the grand-design spiral arms first emerge, and are they long-lived? Future NIRSpec/IFU or ALMA observations will be helpful to ascertain whether the disk of \ourobject is dynamically hot or cold providing constraints on the formation mechanism behind the spiral arms, in this intriguing galaxy.

\begin{acknowledgements}
We acknowledge the support of the Department of Atomic Energy, Government of India, under project no. 12-R\&D-TFR5.02-0700. We thank the referee for their insightful comments that improved both the content and presentation of this paper. We thank Pralay Biswas for useful discussions.
\end{acknowledgements}

\appendix
\section{}
\label{sec:appendix}

\subsection{BAGPIPES SED modeling parameters, fitted model, star formation history, and corner plots}
\begin{figure*}
\centering
  \begin{tikzpicture}
        \node[anchor=south west,inner sep=0] (image) at (0,0) {\includegraphics[width = 0.9\textwidth]{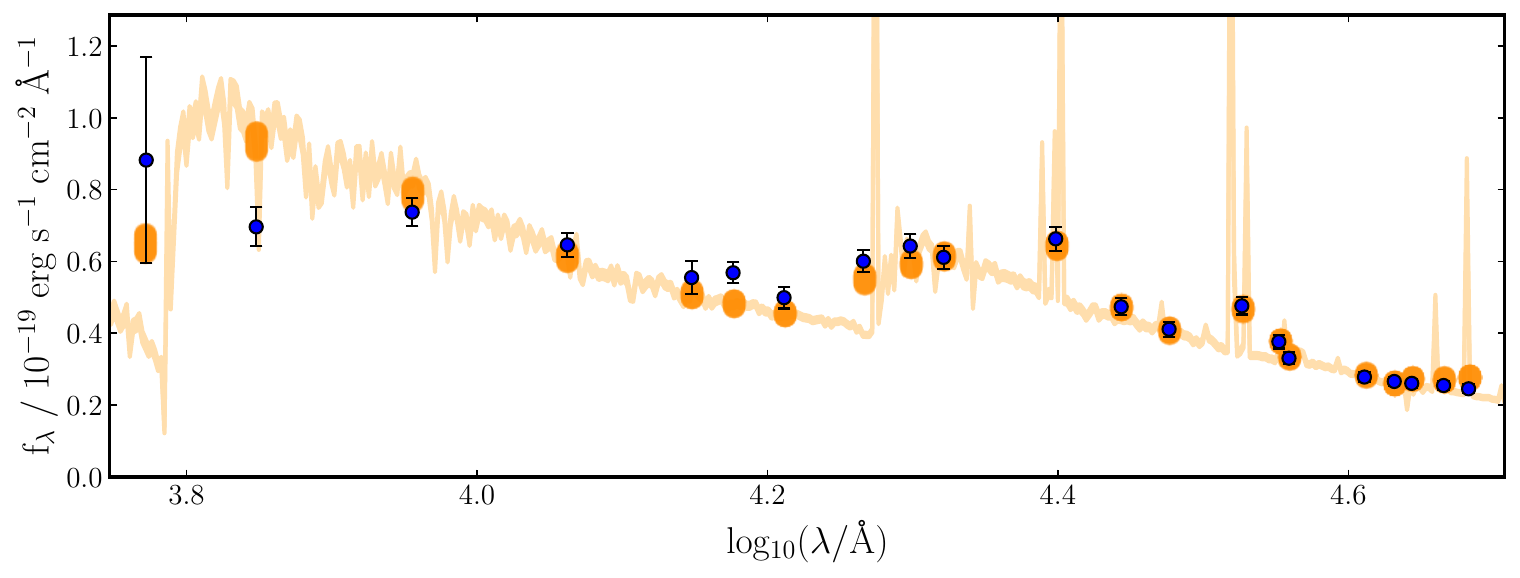}};
        \begin{scope}[x={(image.south east)},y={(image.north west)}]
            \draw[red, thick] (0.108,0.18) -- (0.108,0.23);
            \draw[red, thick] (0.575,0.18) -- (0.575,0.23);
            \draw[red, thick] (0.69,0.18) -- (0.69,0.23);
            \draw[red, thick] (0.815,0.18) -- (0.815,0.23);
        \end{scope}
    \end{tikzpicture}
  \includegraphics[width = 0.9\textwidth]{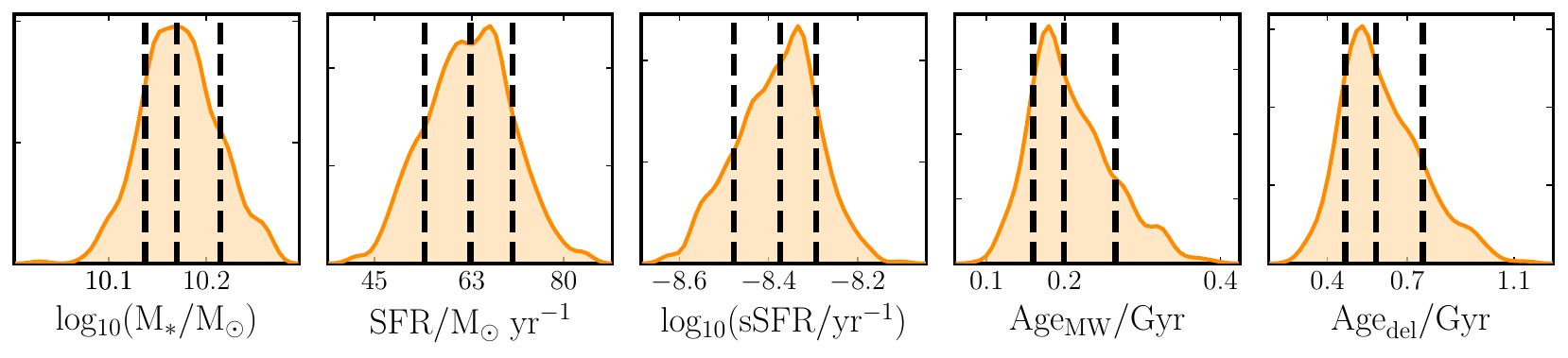}
  \includegraphics[width = 0.9\textwidth]{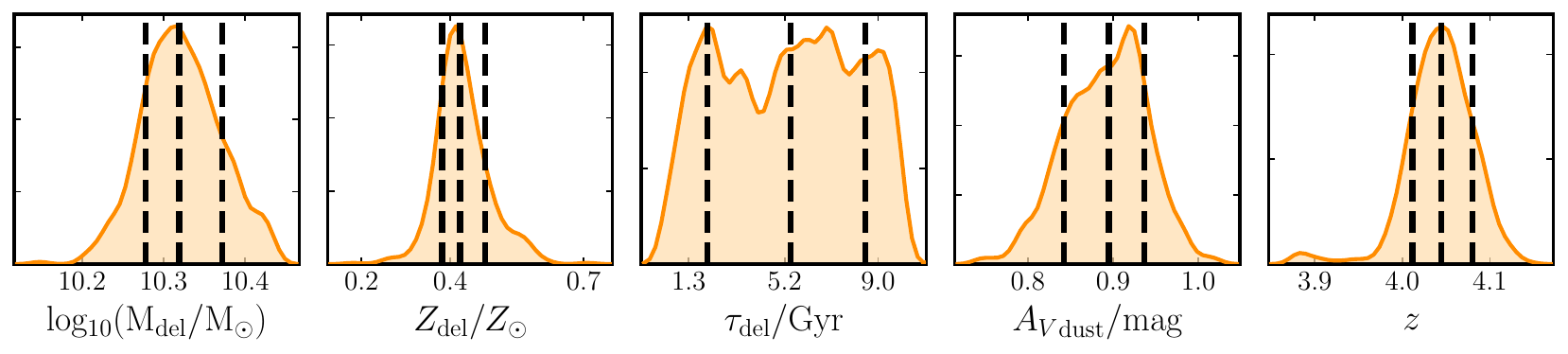}
  \caption{In the top panel, we show the fitted model (as an orange curve), the model fluxes in each filter (as orange points) and the observed photometry with error bars (as blue points) for the delayed SFH model. Along the X-axis, red-tick marks indicate the positions, from left to right, of the Lyman break, the Balmer break, [OIII] + H-$\beta$, and H-$\alpha$ + [NII] line complexes. In the lower panels, we show the 1-d posteriors of output SPS parameters such as stellar mass, SFR, sSFR, age, formed mass (M$_{del}$), metallicity ($\log(Z_{del}/Z_\odot$), $\tau$, dust attenuation (A$_{v}$) and redshift (labeled on the x-axis) with three black vertical dashed lines indicating the values of the 16th, 50th and 84th percentiles of the output posterior probability distribution of each parameter. Age$_{MW}$ is the mass-weighted age of the galaxy, and Age$_{del}$ is the amount of time that has passed since the onset of star formation in the galaxy, calculated from the modeled delayed SFH.} 
  \label{fig:Bagpipes_fit_1dpost}
  
\end{figure*}

\begin{figure*}
\centering
  \includegraphics[width = 0.9\textwidth]{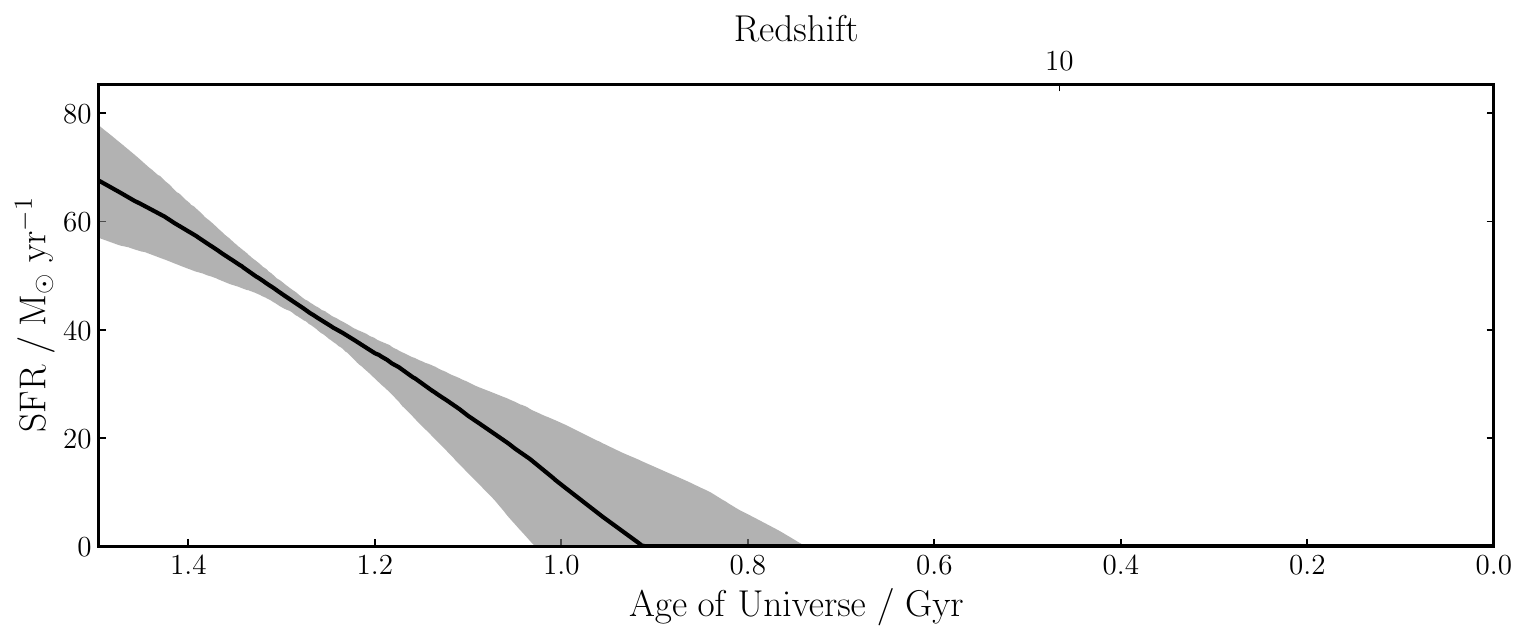}
  \includegraphics[width = 0.9\textwidth]{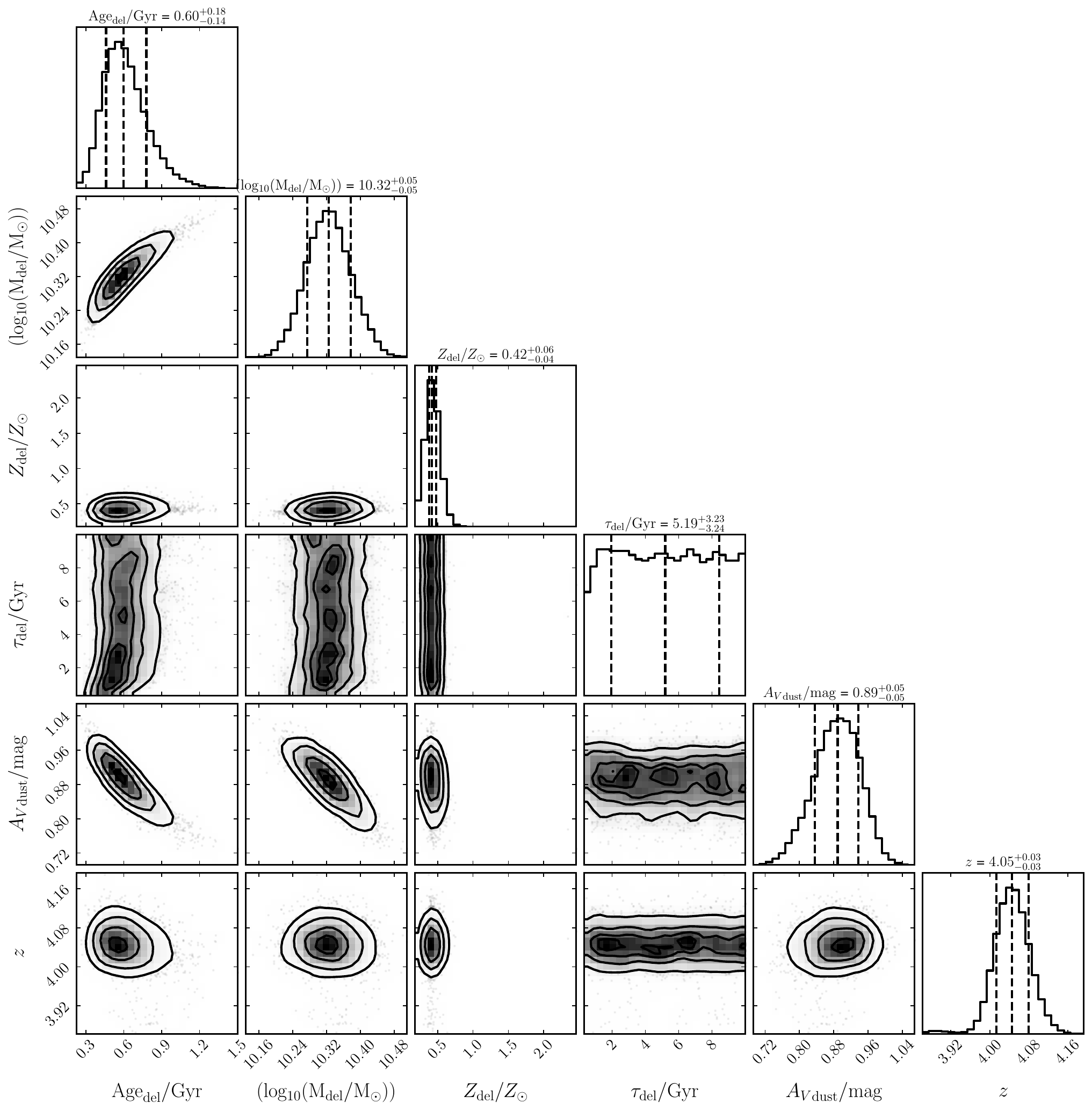}
  \caption{In the upper panel of this figure, we show the delayed exponential star formation history fitted by BAGPIPES to the observed photometry of the galaxy. In the lower panel, we show the corner plots for the output parameters.}
  \label{fig:Bagpipes_corner_sfh}

\end{figure*}

We have used the SED modeling software BAGPIPES to derive joint constraints on the photometric redshift and SPS properties. As input, we provide the ranges of SPS parameters and star formation histories (Section~\ref{sec:methods}). We utilize the photometry as provided by the UNCOVER DR3 catalog fluxes in different filters. For flux errors, we impose a 5\% uncertainty error floor added in quadrature with the flux uncertainties given in the UNCOVER catalog\citep{Boyer_2022}. The range and meaning of the input parameters are provided in Table~\ref{tab:Bagpipes_param}. The rest of the parameters have been kept at their default configuration. We have fitted exponentially decaying and delayed exponentially decaying star formation histories (SFHs). In Fig~\ref{fig:Bagpipes_fit_1dpost} and Fig~\ref{fig:Bagpipes_corner_sfh} we provide the output SED fit, one-dimensional posteriors, output SED fit, fitted SFH, and corner plots obtained for the delayed star formation histories.


\begin{table*}
\caption{\label{tab:Bagpipes_param}Input parameters used for SED modeling with BAGPIPES}

\centering
\begin{tabular}{lcc}
\hline\hline
Parameter & Range & Remarks \\

\hline
Redshift & 0 - 10 &   \\
Stellar mass & 7 - 15 log \(M_\odot\) & log uniform prior \\
Metallicity & 0 - 2.5 \(Z_\odot\)&  \\
Dust ($A_{v}$) & 0 - 4 &Dust attenuation parameter in V-band (Calzetti dust law)  \\
Dust (Eta) & 2 &  Multiplicative factor on Av for stars in birth clouds \citep{Charlot_2000}\\
Nebular emission& $\log U = -3$ &logarithm of the ionization parameter   \\
Age & 0.1 - 15 Gyr &  Restricted to the age of the Universe at the redshift  \\
$\tau$ & 0.3 - 10 Gyr &  Half-life timescale of the decrease of star formation \\

\hline
\end{tabular}
 \end{table*}

\subsection{Best fit Prospector model from UNCOVER DR4 SPS modeling}
\begin{figure*}
\centering
  \includegraphics[width = 0.9\textwidth]{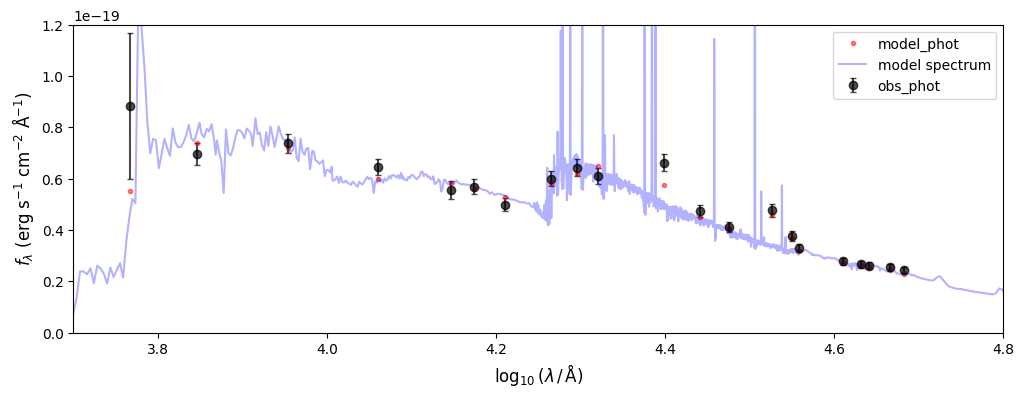}

  \caption{The best fit SED model from Prospector derived from UNCOVER DR4 SPS output files. Observed flux, flux uncertainties and model fluxes in all the filters are plotted. Flux uncertainties are taken from UNCOVER DR4 SPS data release, which have a 5\% error floor imposed on the catalog flux uncertainties due to potential calibration uncertainties with JWST photometry. Wavelength is plotted in log scale in units of \r{A}ngstrom.}
  \label{fig:Pros_best}

\end{figure*}

The Prospector best fit SED model flux estimates are similar to observed fluxes within error bars, except in a few filters. For example, flux in the F250M filter is underestimated in the Prospector SED model (see Fig~\ref{fig:Pros_best}).

\bibliographystyle{aa} 
\bibliography{manuscript} 

\begin{thebibliography}{86}
\expandafter\ifx\csname natexlab\endcsname\relax\def\natexlab#1{#1}\fi

\bibitem[{{Abraham} {et~al.}(1996){Abraham}, {Tanvir}, {Santiago}, {Ellis}, {Glazebrook}, \& {van den Bergh}}]{Abraham_1996}
{Abraham}, R.~G., {Tanvir}, N.~R., {Santiago}, B.~X., {et~al.} 1996, \mnras, 279, L47

\bibitem[{{Abraham} {et~al.}(2003){Abraham}, {van den Bergh}, \& {Nair}}]{Abraham_2003}
{Abraham}, R.~G., {van den Bergh}, S., \& {Nair}, P. 2003, \apj, 588, 218

\bibitem[{{Akhlaghi}(2019{\natexlab{a}})}]{noisechisel_segment_2019}
{Akhlaghi}, M. 2019{\natexlab{a}}, arXiv e-prints, arXiv:1909.11230

\bibitem[{{Akhlaghi}(2019{\natexlab{b}})}]{makecatalog}
{Akhlaghi}, M. 2019{\natexlab{b}}, ASPC, 521, 299

\bibitem[{{Akhlaghi} \& {Ichikawa}(2015)}]{gnuastro}
{Akhlaghi}, M. \& {Ichikawa}, T. 2015, ApJS, 220, 1

\bibitem[{Bershady {et~al.}(2000)Bershady, Jangren, \& Conselice}]{Bershady_2000}
Bershady, M.~A., Jangren, A., \& Conselice, C.~J. 2000, The Astronomical Journal, 119, 2645

\bibitem[{{Bezanson} {et~al.}(2022){Bezanson}, {Labbe}, {Whitaker}, {Leja}, {Price}, {Franx}, {Brammer}, {Marchesini}, {Zitrin}, {Wang}, {Weaver}, {Furtak}, {Atek}, {Coe}, {Cutler}, {Dayal}, {van Dokkum}, {Feldmann}, {Forster Schreiber}, {Fujimoto}, {Geha}, {Glazebrook}, {de Graaff}, {Greene}, {Juneau}, {Kassin}, {Kriek}, {Khullar}, {Maseda}, {Mowla}, {Muzzin}, {Nanayakkara}, {Nelson}, {Oesch}, {Pacifici}, {Pan}, {Papovich}, {Setton}, {Shapley}, {Smit}, {Stefanon}, {Taylor}, \& {Williams}}]{Bezanson_2022}
{Bezanson}, R., {Labbe}, I., {Whitaker}, K.~E., {et~al.} 2022, arXiv e-prints, arXiv:2212.04026

\bibitem[{{Bournaud} \& {Elmegreen}(2009)}]{Bournaud_2009}
{Bournaud}, F. \& {Elmegreen}, B.~G. 2009, \apjl, 694, L158

\bibitem[{Bournaud {et~al.}(2007)Bournaud, Elmegreen, \& Elmegreen}]{Bournaud_2007}
Bournaud, F., Elmegreen, B.~G., \& Elmegreen, D.~M. 2007, The Astrophysical Journal, 670, 237

\bibitem[{Boyer {et~al.}(2022)Boyer, Anderson, Gennaro, Geha, McQuinn, Tollerud, Correnti, Newman, Cohen, Kallivayalil, Beaton, Cole, Dolphin, Kalirai, Sandstrom, Savino, Skillman, Weisz, \& Williams}]{Boyer_2022}
Boyer, M.~L., Anderson, J., Gennaro, M., {et~al.} 2022, Research Notes of the AAS, 6, 191

\bibitem[{{Brammer} {et~al.}(2008){Brammer}, {van Dokkum}, \& {Coppi}}]{Brammer_2008}
{Brammer}, G.~B., {van Dokkum}, P.~G., \& {Coppi}, P. 2008, \apj, 686, 1503

\bibitem[{Bruzual \& Charlot(2003)}]{Bruzal&Charlot_2003}
Bruzual, G. \& Charlot, S. 2003, Monthly Notices of the Royal Astronomical Society, 344, 1000

\bibitem[{Buchner {et~al.}(2014)Buchner, Georgakakis, Nandra, Hsu, Rangel, Brightman, Merloni, Salvato, Donley, \& Kocevski}]{Buchner_2014}
Buchner, J., Georgakakis, A., Nandra, K., {et~al.} 2014, Astronomy \& Astrophysics, 564, A125

\bibitem[{{Calzetti} {et~al.}(2000){Calzetti}, {Armus}, {Bohlin}, {Kinney}, {Koornneef}, \& {Storchi-Bergmann}}]{Calzetti_2000}
{Calzetti}, D., {Armus}, L., {Bohlin}, R.~C., {et~al.} 2000, \apj, 533, 682

\bibitem[{Carnall {et~al.}(2018)Carnall, McLure, Dunlop, \& Davé}]{Carnall_2018}
Carnall, A.~C., McLure, R.~J., Dunlop, J.~S., \& Davé, R. 2018, Monthly Notices of the Royal Astronomical Society, 480, 4379–4401

\bibitem[{{Charlot} \& {Fall}(2000)}]{Charlot_2000}
{Charlot}, S. \& {Fall}, S.~M. 2000, \apj, 539, 718

\bibitem[{{Conroy} \& {Gunn}(2010)}]{Conroy&Gunn_2010}
{Conroy}, C. \& {Gunn}, J.~E. 2010, \apj, 712, 833

\bibitem[{{Conroy} {et~al.}(2010){Conroy}, {White}, \& {Gunn}}]{Conroy_2010}
{Conroy}, C., {White}, M., \& {Gunn}, J.~E. 2010, \apj, 708, 58

\bibitem[{{Conselice}(2003)}]{Conselice_2003}
{Conselice}, C.~J. 2003, \apjs, 147, 1

\bibitem[{{Costantin} {et~al.}(2023){Costantin}, {P{\'e}rez-Gonz{\'a}lez}, {Guo}, {Buttitta}, {Jogee}, {Bagley}, {Barro}, {Kartaltepe}, {Koekemoer}, {Cabello}, {Corsini}, {M{\'e}ndez-Abreu}, {de la Vega}, {Iyer}, {Bisigello}, {Cheng}, {Morelli}, {Arrabal Haro}, {Buitrago}, {Cooper}, {Dekel}, {Dickinson}, {Finkelstein}, {Giavalisco}, {Holwerda}, {Huertas-Company}, {Lucas}, {Papovich}, {Pirzkal}, {Seill{\'e}}, {Vega-Ferrero}, {Wuyts}, \& {Yung}}]{Costantin_2023}
{Costantin}, L., {P{\'e}rez-Gonz{\'a}lez}, P.~G., {Guo}, Y., {et~al.} 2023, \nat, 623, 499

\bibitem[{{Danhaive} {et~al.}(2025){Danhaive}, {Tacchella}, {\textbackslash''Ubler}, {de Graaff}, {Egami}, {Johnson}, {Sun}, {Arribas}, {Bunker}, {Carniani}, {Jones}, {Maiolino}, {McClymont}, {Parlanti}, {Simmonds}, {Villanueva}, {Baker}, {Jaffe}, {Eisenstein}, {Hainline}, {Helton}, {Ji}, {Lin}, {Pusk\textbackslash'as}, {Rieke}, {Rinaldi}, {Robertson}, {Scholz}, {Williams}, \& {Willmer}}]{Danhaive_2025}
{Danhaive}, A.~L., {Tacchella}, S., {\textbackslash''Ubler}, H., {et~al.} 2025, arXiv e-prints, arXiv:2503.21863

\bibitem[{{de Graaff} {et~al.}(2024){de Graaff}, {Rix}, {Carniani}, {Suess}, {Charlot}, {Curtis-Lake}, {Arribas}, {Baker}, {Boyett}, {Bunker}, {Cameron}, {Chevallard}, {Curti}, {Eisenstein}, {Franx}, {Hainline}, {Hausen}, {Ji}, {Johnson}, {Jones}, {Maiolino}, {Maseda}, {Nelson}, {Parlanti}, {Rawle}, {Robertson}, {Tacchella}, {{\"U}bler}, {Williams}, {Willmer}, \& {Willott}}]{Graaff_2024}
{de Graaff}, A., {Rix}, H.-W., {Carniani}, S., {et~al.} 2024, \aap, 684, A87

\bibitem[{{Dobbs} \& {Baba}(2014)}]{Dobbs_2014}
{Dobbs}, C. \& {Baba}, J. 2014, \pasa, 31, e035

\bibitem[{Elmegreen {et~al.}(2009{\natexlab{a}})Elmegreen, Elmegreen, Fernandez, \& Lemonias}]{Elmegreen_2009a}
Elmegreen, B.~G., Elmegreen, D.~M., Fernandez, M.~X., \& Lemonias, J.~J. 2009{\natexlab{a}}, The Astrophysical Journal, 692, 12

\bibitem[{{Elmegreen} \& {Elmegreen}(1982)}]{Elmegreen_1982}
{Elmegreen}, D.~M. \& {Elmegreen}, B.~G. 1982, \mnras, 201, 1021

\bibitem[{{Elmegreen} \& {Elmegreen}(2014)}]{Elmegreen_2014}
{Elmegreen}, D.~M. \& {Elmegreen}, B.~G. 2014, \apj, 781, 11

\bibitem[{Elmegreen {et~al.}(2009{\natexlab{b}})Elmegreen, Elmegreen, Marcus, Shahinyan, Yau, \& Petersen}]{Elmegreen_2009b}
Elmegreen, D.~M., Elmegreen, B.~G., Marcus, M.~T., {et~al.} 2009{\natexlab{b}}, The Astrophysical Journal, 701, 306

\bibitem[{{Elmegreen} {et~al.}(2007){Elmegreen}, {Elmegreen}, {Ravindranath}, \& {Coe}}]{Elmegreen_DM_2007}
{Elmegreen}, D.~M., {Elmegreen}, B.~G., {Ravindranath}, S., \& {Coe}, D.~A. 2007, \apj, 658, 763

\bibitem[{{Feng} {et~al.}(2015){Feng}, {Di Matteo}, {Croft}, {Tenneti}, {Bird}, {Battaglia}, \& {Wilkins}}]{Feng_2015}
{Feng}, Y., {Di Matteo}, T., {Croft}, R., {et~al.} 2015, \apjl, 808, L17

\bibitem[{{Feng} {et~al.}(2016){Feng}, {Di-Matteo}, {Croft}, {Bird}, {Battaglia}, \& {Wilkins}}]{Feng_2016}
{Feng}, Y., {Di-Matteo}, T., {Croft}, R.~A., {et~al.} 2016, \mnras, 455, 2778

\bibitem[{Feroz \& Hobson(2008)}]{Feroz&Hobson_2008}
Feroz, F. \& Hobson, M.~P. 2008, Monthly Notices of the Royal Astronomical Society, 384, 449

\bibitem[{Feroz {et~al.}(2009)Feroz, Hobson, \& Bridges}]{Feroz_2009}
Feroz, F., Hobson, M.~P., \& Bridges, M. 2009, Monthly Notices of the Royal Astronomical Society, 398, 1601

\bibitem[{Feroz {et~al.}(2019)Feroz, Hobson, Cameron, \& Pettitt}]{Feroz_2019}
Feroz, F., Hobson, M.~P., Cameron, E., \& Pettitt, A.~N. 2019, The Open Journal of Astrophysics, 2

\bibitem[{Ferreira {et~al.}(2022)Ferreira, Adams, Conselice, Sazonova, Austin, Caruana, Ferrari, Verma, Trussler, Broadhurst, Diego, Frye, Pascale, Wilkins, Windhorst, \& Zitrin}]{Ferreira_2022}
Ferreira, L., Adams, N., Conselice, C.~J., {et~al.} 2022, The Astrophysical Journal Letters, 938, L2

\bibitem[{{Ferreira} {et~al.}(2023){Ferreira}, {Conselice}, {Sazonova}, {Ferrari}, {Caruana}, {Tohill}, {Lucatelli}, {Adams}, {Irodotou}, {Marshall}, {Roper}, {Lovell}, {Verma}, {Austin}, {Trussler}, \& {Wilkins}}]{Ferreira_2023}
{Ferreira}, L., {Conselice}, C.~J., {Sazonova}, E., {et~al.} 2023, \apj, 955, 94

\bibitem[{{Freeman}(1970)}]{Freeman_1970}
{Freeman}, K.~C. 1970, \apj, 160, 811

\bibitem[{{Fudamoto} {et~al.}(2022){Fudamoto}, {Inoue}, \& {Sugahara}}]{Fudamoto_2022}
{Fudamoto}, Y., {Inoue}, A.~K., \& {Sugahara}, Y. 2022, \apjl, 938, L24

\bibitem[{{Furtak} {et~al.}(2023){Furtak}, {Zitrin}, {Weaver}, {Atek}, {Bezanson}, {Labb{\'e}}, {Whitaker}, {Leja}, {Price}, {Brammer}, {Wang}, {Marchesini}, {Pan}, {Dayal}, {van Dokkum}, {Feldmann}, {Fujimoto}, {Franx}, {Khullar}, {Nelson}, \& {Mowla}}]{Furtak_2023}
{Furtak}, L.~J., {Zitrin}, A., {Weaver}, J.~R., {et~al.} 2023, \mnras, 523, 4568

\bibitem[{{Gardner} {et~al.}(2023){Gardner}, {Mather}, {Abbott}, {Abell}, {Abernathy}, {Abney}, {Abraham}, {Abraham}, {Abul-Huda}, {Acton}, {Adams}, {Adams}, {Adler}, {Adriaensen}, {Aguilar}, {Ahmed}, {Ahmed}, {Ahmed}, {Albat}, {Albert}, {Alberts}, {Aldridge}, {Allen}, {Allen}, {Altenburg}, {Altunc}, {Alvarez}, {{\'A}lvarez-M{\'a}rquez}, {Alves de Oliveira}, {Ambrose}, {Anandakrishnan}, {Andersen}, {Anderson}, {Anderson}, {Anderson}, {Anderson}, {Aprea}, {Archer}, {Arenberg}, {Argyriou}, {Arribas}, {Artigau}, {Arvai}, {Atcheson}, {Atkinson}, {Averbukh}, {Aymergen}, {Bacinski}, {Baggett}, {Bagnasco}, {Baker}, {Balzano}, {Banks}, {Baran}, {Barker}, {Barrett}, {Barringer}, {Barto}, {Bast}, {Baudoz}, {Baum}, {Beatty}, {Beaulieu}, {Bechtold}, {Beck}, {Beddard}, {Beichman}, {Bellagama}, {Bely}, {Berger}, {Bergeron}, {Bernier}, {Bertch}, {Beskow}, {Betz}, {Biagetti}, {Birkmann}, {Bjorklund}, {Blackwood}, {Blazek}, {Blossfeld}, {Bluth}, {Boccaletti}, {Boegner}, {Bohlin}, {Boia}, {B{\"o}ker}, {Bonaventura}, {Bond},
  {Bosley}, {Boucarut}, {Bouchet}, {Bouwman}, {Bower}, {Bowers}, {Bowers}, {Boyce}, {Boyer}, {Boyer}, {Boyer}, {Boyer}, {Bradley}, {Brady}, {Brandl}, {Brannen}, {Breda}, {Bremmer}, {Brennan}, {Bresnahan}, {Bright}, {Broiles}, {Bromenschenkel}, {Brooks}, {Brooks}, {Brown}, {Brown}, {Brown}, {Bruce}, {Bryson}, {Bujanda}, {Bullock}, {Bunker}, {Bureo}, {Burt}, {Bush}, {Bushouse}, {Bussman}, {Cabaud}, {Cale}, {Calhoon}, {Calvani}, {Canipe}, {Caputo}, {Cara}, {Carey}, {Case}, {Cesari}, {Cetorelli}, {Chance}, {Chandler}, {Chaney}, {Chapman}, {Charlot}, {Chayer}, {Cheezum}, {Chen}, {Chen}, {Cherinka}, {Chichester}, {Chilton}, {Chittiraibalan}, {Clampin}, {Clark}, {Clark}, {Clark}, {Claybrooks}, {Cleveland}, {Cohen}, {Cohen}, {Col{\'o}n}, {Coleman}, {Colina}, {Comber}, {Comeau}, {Comer}, {Conde Reis}, {Connolly}, {Conroy}, {Contos}, {Contreras}, {Cook}, {Cooper}, {Cooper}, {Correia}, {Correnti}, {Cossou}, {Costanza}, {Coulais}, {Cox}, {Coyle}, {Cracraft}, {Crew}, {Curtis}, {Cusveller}, {Da Costa Maciel}, {Dailey},
  {Daugeron}, {Davidson}, {Davies}, {Davis}, {Davis}, {Day}, {de Chambure}, {de Jong}, {De Marchi}, {Dean}, {Decker}, {Delisa}, {Dell}, \& {Dellagatta}}]{Gardner_2023}
{Gardner}, J.~P., {Mather}, J.~C., {Abbott}, R., {et~al.} 2023, \pasp, 135, 068001

\bibitem[{{Genzel} {et~al.}(2006){Genzel}, {Tacconi}, {Eisenhauer}, {F{\"o}rster Schreiber}, {Cimatti}, {Daddi}, {Bouch{\'e}}, {Davies}, {Lehnert}, {Lutz}, {Nesvadba}, {Verma}, {Abuter}, {Shapiro}, {Sternberg}, {Renzini}, {Kong}, {Arimoto}, \& {Mignoli}}]{Genzel_2006}
{Genzel}, R., {Tacconi}, L.~J., {Eisenhauer}, F., {et~al.} 2006, \nat, 442, 786

\bibitem[{{Guo} {et~al.}(2015){Guo}, {Ferguson}, {Bell}, {Koo}, {Conselice}, {Giavalisco}, {Kassin}, {Lu}, {Lucas}, {Mandelker}, {McIntosh}, {Primack}, {Ravindranath}, {Barro}, {Ceverino}, {Dekel}, {Faber}, {Fang}, {Koekemoer}, {Noeske}, {Rafelski}, \& {Straughn}}]{Guo_2015}
{Guo}, Y., {Ferguson}, H.~C., {Bell}, E.~F., {et~al.} 2015, \apj, 800, 39

\bibitem[{{Hernquist}(1992)}]{Hernquist_1992}
{Hernquist}, L. 1992, \apj, 400, 460

\bibitem[{{Hopkins} {et~al.}(2008){Hopkins}, {Hernquist}, {Cox}, \& {Kere{\v{s}}}}]{Hopkins_2008}
{Hopkins}, P.~F., {Hernquist}, L., {Cox}, T.~J., \& {Kere{\v{s}}}, D. 2008, \apjs, 175, 356

\bibitem[{{Huang} {et~al.}(2023){Huang}, {Li}, {Cheng}, {Hou}, {Yan}, {Willner}, {Dai}, {Zheng}, {Pan}, {Rigopoulou}, {Wang}, {Li}, {Liang}, {Esamdin}, \& {Fazio}}]{Huang_J_2023}
{Huang}, J.~S., {Li}, Z.-J., {Cheng}, C., {et~al.} 2023, \apj, 949, 83

\bibitem[{Huang {et~al.}(2023)Huang, Kawabe, Kohno, Saito, Mizukoshi, Iono, Michiyama, Tamura, Hayward, \& Umehata}]{Huang_2023}
Huang, S., Kawabe, R., Kohno, K., {et~al.} 2023, The Astrophysical Journal Letters, 958, L26

\bibitem[{{Jacobs} {et~al.}(2023){Jacobs}, {Glazebrook}, {Calabr{\`o}}, {Treu}, {Nannayakkara}, {Jones}, {Merlin}, {Abraham}, {Stevens}, {Vulcani}, {Yang}, {Bonchi}, {Boyett}, {Brada{\v{c}}}, {Castellano}, {Fontana}, {Marchesini}, {Malkan}, {Mason}, {Morishita}, {Paris}, {Santini}, {Trenti}, \& {Wang}}]{Jacobs_2023}
{Jacobs}, C., {Glazebrook}, K., {Calabr{\`o}}, A., {et~al.} 2023, \apjl, 948, L13

\bibitem[{{Johnson} {et~al.}(2021){Johnson}, {Leja}, {Conroy}, \& {Speagle}}]{Johnson_2021}
{Johnson}, B.~D., {Leja}, J., {Conroy}, C., \& {Speagle}, J.~S. 2021, \apjs, 254, 22

\bibitem[{Kartaltepe {et~al.}(2023)Kartaltepe, Rose, Vanderhoof, McGrath, Costantin, Cox, Yung, Kocevski, Wuyts, Ferguson, Bagley, Finkelstein, Amorín, Andrews, Haro, Backhaus, Behroozi, Bisigello, Calabrò, Casey, Coogan, Cooper, Croton, de~la Vega, Dickinson, Fontana, Franco, Grazian, Grogin, Hathi, Holwerda, Huertas-Company, Iyer, Jogee, Jung, Kewley, Kirkpatrick, Koekemoer, Liu, Lotz, Lucas, Newman, Pacifici, Pandya, Papovich, Pentericci, Pérez-González, Petersen, Pirzkal, Rafelski, Ravindranath, Simons, Snyder, Somerville, Stanway, Straughn, Tacchella, Trump, Vega-Ferrero, Wilkins, Yang, \& Zavala}]{Kartaltepe_2023}
Kartaltepe, J.~S., Rose, C., Vanderhoof, B.~N., {et~al.} 2023, The Astrophysical Journal Letters, 946, L15

\bibitem[{{Kron}(1980)}]{Kron_1980}
{Kron}, R.~G. 1980, \apjs, 43, 305

\bibitem[{{Kroupa} \& {Boily}(2002)}]{Kroupa_2002}
{Kroupa}, P. \& {Boily}, C.~M. 2002, \mnras, 336, 1188

\bibitem[{Kuhn {et~al.}(2024)Kuhn, Guo, Martin, Bayless, Gates, \& Puleo}]{Kuhn_2024}
Kuhn, V., Guo, Y., Martin, A., {et~al.} 2024, The Astrophysical Journal Letters, 968, L15

\bibitem[{Law {et~al.}(2012)Law, Shapley, Steidel, Reddy, Christensen, \& Erb}]{Law_2012}
Law, D.~R., Shapley, A.~E., Steidel, C.~C., {et~al.} 2012, Nature, 487, 338–340

\bibitem[{{Lin} \& {Shu}(1964)}]{Lin_Shu_1964}
{Lin}, C.~C. \& {Shu}, F.~H. 1964, \apj, 140, 646

\bibitem[{{Lotz} {et~al.}(2008){Lotz}, {Davis}, {Faber}, {Guhathakurta}, {Gwyn}, {Huang}, {Koo}, {Le Floc'h}, {Lin}, {Newman}, {Noeske}, {Papovich}, {Willmer}, {Coil}, {Conselice}, {Cooper}, {Hopkins}, {Metevier}, {Primack}, {Rieke}, \& {Weiner}}]{Lotz_2008b}
{Lotz}, J.~M., {Davis}, M., {Faber}, S.~M., {et~al.} 2008, \apj, 672, 177

\bibitem[{{Lotz} {et~al.}(2004){Lotz}, {Primack}, \& {Madau}}]{Lotz_2004}
{Lotz}, J.~M., {Primack}, J., \& {Madau}, P. 2004, \aj, 128, 163

\bibitem[{Margalef-Bentabol {et~al.}(2022)Margalef-Bentabol, Conselice, Haeussler, Casteels, Lintott, Masters, \& Simmons}]{Margalef_Bentabol_2022}
Margalef-Bentabol, B., Conselice, C.~J., Haeussler, B., {et~al.} 2022, Monthly Notices of the Royal Astronomical Society, 511, 1502

\bibitem[{{Naidu} {et~al.}(2024){Naidu}, {Matthee}, {Kramarenko}, {Weibel}, {Brammer}, {Oesch}, {Lechner}, {Furtak}, {Di Cesare}, {Torralba}, {Kotiwale}, {Bezanson}, {Bouwens}, {Chandra}, {Claeyssens}, {Danhaive}, {Frebel}, {de Graaff}, {Greene}, {Heintz}, {Ji}, {Kashino}, {Katz}, {Labbe}, {Leja}, {Li}, {Maseda}, {Richard}, {Shivaei}, {Simcoe}, {Sobral}, {Suess}, {Tacchella}, \& {Williams}}]{Naidu_2024}
{Naidu}, R.~P., {Matthee}, J., {Kramarenko}, I., {et~al.} 2024, arXiv e-prints, arXiv:2410.01874

\bibitem[{{Nelson} {et~al.}(2023){Nelson}, {Suess}, {Bezanson}, {Price}, {van Dokkum}, {Leja}, {Wang}, {Whitaker}, {Labb{\'e}}, {Barrufet}, {Brammer}, {Eisenstein}, {Gibson}, {Hartley}, {Johnson}, {Heintz}, {Mathews}, {Miller}, {Oesch}, {Sandles}, {Setton}, {Speagle}, {Tacchella}, {Tadaki}, {{\"U}bler}, \& {Weaver}}]{Nelson_2023}
{Nelson}, E.~J., {Suess}, K.~A., {Bezanson}, R., {et~al.} 2023, \apjl, 948, L18

\bibitem[{{Oh} {et~al.}(2008){Oh}, {Kim}, {Lee}, \& {Kim}}]{Oh_2008}
{Oh}, S.~H., {Kim}, W.-T., {Lee}, H.~M., \& {Kim}, J. 2008, \apj, 683, 94

\bibitem[{{Pacifici} {et~al.}(2023){Pacifici}, {Iyer}, {Mobasher}, {da Cunha}, {Acquaviva}, {Burgarella}, {Calistro Rivera}, {Carnall}, {Chang}, {Chartab}, {Cooke}, {Fairhurst}, {Kartaltepe}, {Leja}, {Ma{\l}ek}, {Salmon}, {Torelli}, {Vidal-Garc{\'\i}a}, {Boquien}, {Brammer}, {Brown}, {Capak}, {Chevallard}, {Circosta}, {Croton}, {Davidzon}, {Dickinson}, {Duncan}, {Faber}, {Ferguson}, {Fontana}, {Guo}, {Haeussler}, {Hemmati}, {Jafariyazani}, {Kassin}, {Larson}, {Lee}, {Mantha}, {Marchi}, {Nayyeri}, {Newman}, {Pandya}, {Pforr}, {Reddy}, {Sanders}, {Shah}, {Shahidi}, {Stevans}, {Triani}, {Tyler}, {Vanderhoof}, {de la Vega}, {Wang}, \& {Weston}}]{Pacifici_2023}
{Pacifici}, C., {Iyer}, K.~G., {Mobasher}, B., {et~al.} 2023, \apj, 944, 141

\bibitem[{{Peng} {et~al.}(2002){Peng}, {Ho}, {Impey}, \& {Rix}}]{Peng_2002}
{Peng}, C.~Y., {Ho}, L.~C., {Impey}, C.~D., \& {Rix}, H.-W. 2002, \aj, 124, 266

\bibitem[{{Peng} {et~al.}(2010){Peng}, {Ho}, {Impey}, \& {Rix}}]{Peng_2010}
{Peng}, C.~Y., {Ho}, L.~C., {Impey}, C.~D., \& {Rix}, H.-W. 2010, \aj, 139, 2097

\bibitem[{{Rawat} {et~al.}(2009){Rawat}, {Wadadekar}, \& {De Mello}}]{Rawat_2009}
{Rawat}, A., {Wadadekar}, Y., \& {De Mello}, D. 2009, \apj, 695, 1315

\bibitem[{{Rieke} {et~al.}(2023){Rieke}, {Kelly}, {Misselt}, {Stansberry}, {Boyer}, {Beatty}, {Egami}, {Florian}, {Greene}, {Hainline}, {Leisenring}, {Roellig}, {Schlawin}, {Sun}, {Tinnin}, {Williams}, {Willmer}, {Wilson}, {Clark}, {Rohrbach}, {Brooks}, {Canipe}, {Correnti}, {DiFelice}, {Gennaro}, {Girard}, {Hartig}, {Hilbert}, {Koekemoer}, {Nikolov}, {Pirzkal}, {Rest}, {Robberto}, {Sunnquist}, {Telfer}, {Wu}, {Ferry}, {Lewis}, {Baum}, {Beichman}, {Doyon}, {Dressler}, {Eisenstein}, {Ferrarese}, {Hodapp}, {Horner}, {Jaffe}, {Johnstone}, {Krist}, {Martin}, {McCarthy}, {Meyer}, {Rieke}, {Trauger}, \& {Young}}]{Rieke_2023}
{Rieke}, M.~J., {Kelly}, D.~M., {Misselt}, K., {et~al.} 2023, \pasp, 135, 028001

\bibitem[{{Robertson} {et~al.}(2006){Robertson}, {Bullock}, {Cox}, {Di Matteo}, {Hernquist}, {Springel}, \& {Yoshida}}]{Robertson_2006}
{Robertson}, B., {Bullock}, J.~S., {Cox}, T.~J., {et~al.} 2006, \apj, 645, 986

\bibitem[{{Robertson} {et~al.}(2023){Robertson}, {Tacchella}, {Johnson}, {Hausen}, {Alabi}, {Boyett}, {Bunker}, {Carniani}, {Egami}, {Eisenstein}, {Hainline}, {Helton}, {Ji}, {Kumari}, {Lyu}, {Maiolino}, {Nelson}, {Rieke}, {Shivaei}, {Sun}, {{\"U}bler}, {Williams}, {Willmer}, \& {Witstok}}]{Robertson_2023}
{Robertson}, B.~E., {Tacchella}, S., {Johnson}, B.~D., {et~al.} 2023, \apjl, 942, L42

\bibitem[{{Rodriguez-Gomez} {et~al.}(2019){Rodriguez-Gomez}, {Snyder}, {Lotz}, {Nelson}, {Pillepich}, {Springel}, {Genel}, {Weinberger}, {Tacchella}, {Pakmor}, {Torrey}, {Marinacci}, {Vogelsberger}, {Hernquist}, \& {Thilker}}]{Rodriguez_Gomez_2019}
{Rodriguez-Gomez}, V., {Snyder}, G.~F., {Lotz}, J.~M., {et~al.} 2019, \mnras, 483, 4140

\bibitem[{{Rowland} {et~al.}(2024){Rowland}, {Hodge}, {Bouwens}, {Mancera Pi{\~n}a}, {Hygate}, {Algera}, {Aravena}, {Bowler}, {da Cunha}, {Dayal}, {Ferrara}, {Herard-Demanche}, {Inami}, {van Leeuwen}, {de Looze}, {Oesch}, {Pallottini}, {Phillips}, {Rybak}, {Schouws}, {Smit}, {Sommovigo}, {Stefanon}, \& {van der Werf}}]{Rowland_2024}
{Rowland}, L.~E., {Hodge}, J., {Bouwens}, R., {et~al.} 2024, arXiv e-prints, arXiv:2405.06025

\bibitem[{{Salo} {et~al.}(2010){Salo}, {Laurikainen}, {Buta}, \& {Knapen}}]{Salo_2010}
{Salo}, H., {Laurikainen}, E., {Buta}, R., \& {Knapen}, J.~H. 2010, \apjl, 715, L56

\bibitem[{{Sersic}(1968)}]{Sersic_1968}
{Sersic}, J.~L. 1968, {Atlas de Galaxias Australes}

\bibitem[{{Springel} \& {Hernquist}(2005)}]{Springel_2005}
{Springel}, V. \& {Hernquist}, L. 2005, \apjl, 622, L9

\bibitem[{{Struck} {et~al.}(2011){Struck}, {Dobbs}, \& {Hwang}}]{Struck_2011}
{Struck}, C., {Dobbs}, C.~L., \& {Hwang}, J.-S. 2011, \mnras, 414, 2498

\bibitem[{{Suess} {et~al.}(2024){Suess}, {Weaver}, {Price}, {Pan}, {Wang}, {Bezanson}, {Brammer}, {Cutler}, {Labbe}, {Leja}, {Williams}, {Whitaker}, {Dayal}, {de Graaff}, {Feldmann}, {Franx}, {Fudamoto}, {Fujimoto}, {Furtak}, {Goulding}, {Greene}, {Khullar}, {Kokorev}, {Kriek}, {Lorenz}, {Marchesini}, {Maseda}, {Matthee}, {Miller}, {Mitsuhashi}, {Mowla}, {Muzzin}, {Naidu}, {Nanayakkara}, {Nelson}, {Oesch}, {Setton}, {Shipley}, {Smit}, {Spilker}, {van Dokkum}, \& {Zitrin}}]{Suess_2024}
{Suess}, K.~A., {Weaver}, J.~R., {Price}, S.~H., {et~al.} 2024, arXiv e-prints, arXiv:2404.13132

\bibitem[{{Tacchella} {et~al.}(2024){Tacchella}, {McClymont}, {Scholtz}, {Maiolino}, {Ji}, {Villanueva}, {Charlot}, {D'Eugenio}, {Helton}, {Williams}, {Witstok}, {Bhatawdekar}, {Carniani}, {Chevallard}, {Curti}, {Hainline}, {Ji}, {Johnson}, {Leja}, {Li}, {Maseda}, {Pusk{\'a}s}, {Rieke}, {Robertson}, {Shivaei}, {Silcock}, {Simmonds}, {{\"U}bler}, {Willmer}, \& {Willott}}]{Tacchella_2024}
{Tacchella}, S., {McClymont}, W., {Scholtz}, J., {et~al.} 2024, arXiv e-prints, arXiv:2404.02194

\bibitem[{{Toomre}(1977)}]{Toomre_1977}
{Toomre}, A. 1977, \araa, 15, 437

\bibitem[{{Toomre}(1981)}]{Toomre_1981}
{Toomre}, A. 1981, in Structure and Evolution of Normal Galaxies, ed. S.~M. {Fall} \& D.~{Lynden-Bell}, 111--136

\bibitem[{{Toomre} \& {Toomre}(1972)}]{Toomre_1972}
{Toomre}, A. \& {Toomre}, J. 1972, \apj, 178, 623

\bibitem[{{Tsukui} \& {Iguchi}(2021)}]{Tsukui_2021}
{Tsukui}, T. \& {Iguchi}, S. 2021, Science, 372, 1201

\bibitem[{{Tsukui} {et~al.}(2024){Tsukui}, {Wisnioski}, {Bland-Hawthorn}, {Mai}, {Iguchi}, {Baba}, \& {Freeman}}]{Tsukui_2024}
{Tsukui}, T., {Wisnioski}, E., {Bland-Hawthorn}, J., {et~al.} 2024, \mnras, 527, 8941

\bibitem[{{Tsukui} {et~al.}(2023){Tsukui}, {Wisnioski}, {Krumholz}, \& {Battisti}}]{Tsukui_2023b}
{Tsukui}, T., {Wisnioski}, E., {Krumholz}, M.~R., \& {Battisti}, A. 2023, \mnras, 523, 4654

\bibitem[{{Wang} {et~al.}(2024){Wang}, {Leja}, {Labb{\'e}}, {Bezanson}, {Whitaker}, {Brammer}, {Furtak}, {Weaver}, {Price}, {Zitrin}, {Atek}, {Coe}, {Cutler}, {Dayal}, {van Dokkum}, {Feldmann}, {Marchesini}, {Franx}, {F{\"o}rster Schreiber}, {Fujimoto}, {Geha}, {Glazebrook}, {de Graaff}, {Greene}, {Juneau}, {Kassin}, {Kriek}, {Khullar}, {Maseda}, {Mowla}, {Muzzin}, {Nanayakkara}, {Nelson}, {Oesch}, {Pacifici}, {Pan}, {Papovich}, {Setton}, {Shapley}, {Smit}, {Stefanon}, {Suess}, {Taylor}, \& {Williams}}]{Wang_2024}
{Wang}, B., {Leja}, J., {Labb{\'e}}, I., {et~al.} 2024, \apjs, 270, 12

\bibitem[{{Wang} {et~al.}(2025){Wang}, {Cantalupo}, {Pensabene}, {Galbiati}, {Travascio}, {Steidel}, {Maseda}, {Pezzulli}, {de Beer}, {Fossati}, {Fumagalli}, {Gallego}, {Lazeyras}, {Mackenzie}, {Matthee}, {Nanayakkara}, \& {Quadri}}]{Wang_2025}
{Wang}, W., {Cantalupo}, S., {Pensabene}, A., {et~al.} 2025, Nature Astronomy, 9, 710

\bibitem[{{Weaver} {et~al.}(2024){Weaver}, {Cutler}, {Pan}, {Whitaker}, {Labb{\'e}}, {Price}, {Bezanson}, {Brammer}, {Marchesini}, {Leja}, {Wang}, {Furtak}, {Zitrin}, {Atek}, {Chemerynska}, {Coe}, {Dayal}, {van Dokkum}, {Feldmann}, {F{\"o}rster Schreiber}, {Franx}, {Fujimoto}, {Fudamoto}, {Glazebrook}, {de Graaff}, {Greene}, {Juneau}, {Kassin}, {Kriek}, {Khullar}, {Maseda}, {Mowla}, {Muzzin}, {Nanayakkara}, {Nelson}, {Oesch}, {Pacifici}, {Papovich}, {Setton}, {Shapley}, {Shipley}, {Smit}, {Stefanon}, {Taylor}, {Weibel}, \& {Williams}}]{Weaver_2024}
{Weaver}, J.~R., {Cutler}, S.~E., {Pan}, R., {et~al.} 2024, \apjs, 270, 7

\bibitem[{Wu {et~al.}(2022)Wu, Cai, Sun, Bian, Lin, Li, Li, Bauer, Egami, Fan, González-López, Li, Wang, Yang, Zhang, \& Zou}]{Wu_2022}
Wu, Y., Cai, Z., Sun, F., {et~al.} 2022, The Astrophysical Journal Letters, 942, L1

\bibitem[{{Xiao} {et~al.}(2024){Xiao}, {Williams}, {Oesch}, {Elbaz}, {Dessauges-Zavadsky}, {Marques-Chaves}, {Bing}, {Ji}, {Weibel}, {Bezanson}, {Brammer}, {Casey}, {Cloonan}, {Daddi}, {Dayal}, {Faisst}, {Franx}, {Glazebrook}, {Hutter}, {Kartaltepe}, {Labbe}, {Lagache}, {Lim}, {Magnelli}, {Martinez}, {Maseda}, {Nanayakkara}, {Schaerer}, \& {Whitaker}}]{Xiao_2024}
{Xiao}, M., {Williams}, C.~C., {Oesch}, P.~A., {et~al.} 2024, arXiv e-prints, arXiv:2412.13264

\bibitem[{{Yuan} {et~al.}(2017){Yuan}, {Richard}, {Gupta}, {Federrath}, {Sharma}, {Groves}, {Kewley}, {Cen}, {Birnboim}, \& {Fisher}}]{Yuan_2017}
{Yuan}, T., {Richard}, J., {Gupta}, A., {et~al.} 2017, \apj, 850, 61

\end{thebibliography}

\end{document}